\documentclass{article}
\usepackage{microtype}
\usepackage{graphicx}
\usepackage{array}
\usepackage{longtable}

\PassOptionsToPackage{numbers,sort&compress}{natbib}

\usepackage[preprint]{neurips_2026}

\usepackage[utf8]{inputenc} 
\usepackage[T1]{fontenc}    
\usepackage{hyperref}       
\usepackage{url}            
\usepackage{booktabs}       
\usepackage{amsfonts}       
\usepackage{nicefrac}       
\usepackage{microtype}      
\usepackage{xcolor}         

\title{Towards fNIRS foundation models: A curated corpus of 67 public datasets with channel positions in MNI space}

\author{%
  Santiago~Posso-Murillo \\
  Department of Electrical and Computer Engineering\\
  University of Kentucky\\
  Lexington, KY 40506, US \\
  \texttt{spo230@uky.edu} \\
}

\begin{document}

\maketitle

\begin{abstract}
Brain foundation models have transformed medical artificial intelligence and computational neuroscience by reducing the reliance on task-specific data through large-scale pretraining. While they have been successful with non-invasive neuroimaging data, their adoption in functional near-infrared spectroscopy (fNIRS) has lagged due to the limited size and heterogeneity of public datasets, which are collected under varying protocols, devices, and montages. In addition, no single fNIRS dataset is comparable in size to the largest datasets in functional magnetic resonance imaging (fMRI), electroencephalography (EEG), or magnetoencephalography (MEG). The fNIRS datasets are also spread across different platforms and often have insufficient information, or lack the protocols that would make them easier to use. Given the advances in EEG foundation models that enable flexible modeling across diverse EEG montages, we curated 1,955 hours of fNIRS data from 67 public datasets spanning 2,798 participants, representing the largest fNIRS curation effort to date. The curated data include oxygenated (HbO) and deoxygenated (HbR) hemoglobin signals along with the three-dimensional position of every channel in the standard space of the Montreal Neurological Institute (MNI), which makes positions comparable across datasets. We release dataloaders that handle each dataset automatically, along with the minimal preprocessing pipeline used to prepare the data, at \url{https://github.com/sposso/fnirs-datasets}. This work aims to facilitate the future development of foundation models for fNIRS.
\end{abstract}

\section{Introduction}

fNIRS is a noninvasive neuroimaging technique that measures changes in cortical HbO and HbR, a hemodynamic proxy for neural activity~\cite{Scholkmann2014}. Compared with fMRI, fNIRS is inexpensive, portable and has a higher temporal resolution. Compared with EEG, it is more robust to motion artifacts~\cite{pinti2020present}, which allows measurements during movement and in naturalistic settings~\cite{yucel2021best}. Due to its practical properties fNIRS has been applied to stroke rehabilitation, Alzheimer's screening, pediatric developmental disorders, neonatal brain injury, psychiatric assessment, and the decoding of speech and movement~\cite{Keles2022,Conti2022,Shin2023,chao2021fnirs,Tran2023,PossoMurillo2025,posso2026towards,Li2025}.

Over the years, we have witnessed the growth of foundation models across various domains that outperform traditional machine learning methods. A foundation model is defined as any model that is trained on broad data and learns generalizable representations that can be adapted to a wide range of downstream tasks \cite{bommasani2021opportunities}. This approach has driven groundbreaking progress in natural language processing \cite{zhou2025comprehensive}, computer vision~\cite{awais2025foundation}, medical imaging~\cite{Moor2023,Ryu2025}, and recently in noninvasive neuroimaging techniques.

In EEG, several foundation models have been proposed~\cite{kostas2021bendr,Yang2023,jiang2024large,doner2026luna} and fine-tuned for seizure detection, emotion recognition, sleep staging and abnormality detection. Comparable models have been reported for fMRI \cite{ortega2024brainlm,dong2024brain,xia2026brain}, including one pretrained on more than 50,000 subjects that predicted 17 diagnoses across hospital cohorts from two countries~\cite{Wang2026}. Similar progress has followed in MEG~\cite{huang2025meg,liao2026pretrained}. However, foundation models for fNIRS remain underexplored for several reasons.

First, public fNIRS datasets are heterogeneous. They differ in device, number of channels, probe layout and cortical coverage. Additionally, they are acquired under diverse experimental paradigms, and many report the optode positions incompletely or not at all. Second, they are small. Of the more than 90 public datasets reviewed in this work, the median holds 18~h of recordings and the largest 265~h, whereas single EEG and fMRI datasets used to pretrain foundation models exceed 21,000~h~\cite{Obeid2016} and 6,700~h~\cite{miller2016multimodal}, respectively. Third, they are scattered across repositories, in different formats and with different levels of documentation.

Despite these limitations, a few fNIRS foundation models have been proposed. EFRM, a multimodal EEG--fNIRS model, was pretrained on about 1,250~h of recordings, of which 364~h were fNIRS~\cite{jung2025efrm}. Subedi et al.\ pretrained a model on recordings from one procedural setting with one probe layout~\cite{subedi2026interpretable}. GeoNIRS, the fNIRS foundation model pretrained on the most data to date, used 827~h from 22 datasets~\cite{li2026geonirs}, less than 5\% of the hours used to pretrain some EEG models \cite{doner2026luna}.

To support the development of fNIRS foundation models, we reviewed more than 90 public fNIRS datasets and organized them in three groups. The first group is a pretraining corpus of 61 datasets, with 2,577 participants and 1,836~h of recordings, more than twice the largest pretraining set used for fNIRS to date~\cite{li2026geonirs}. The recordings come from laboratories in Asia, Europe, and North America and were acquired with continuous-wave instruments from diverse manufacturers. The probes range from sparse commercial arrays of a few channels to high-density diffuse optical tomography systems with hundreds of channels. The participants include infants, children and adults, healthy or with neurological conditions such as stroke, mild cognitive impairment (MCI), Parkinson's disease (PD) and Alzheimer's disease (AD). The paradigms cover seven categories: resting state, motor, cognitive, speech, visual, emotion and social interaction (Table~\ref{tab:summary}). The second group holds six datasets held out for evaluation with 221 participants and 119~h of recordings (Table~\ref{tab:downstream}), taken from the downstream benchmark of GeoNIRS~\cite{li2026geonirs}. This allows future fNIRS foundation models to be evaluated on the same tasks.

All 67 datasets provide HbO and HbR signals and 3D optode positions placed in MNI space, which makes positions comparable across datasets (Section~\ref{CS}). The position criterion is needed to the lack of a convention for naming fNIRS optodes. EEG electrodes carry the standardized names of the 10--20 system, which foundation models have used as a prior to learn the spatial information of each channel~\cite{Yang2023,jiang2024large}. However, fNIRS channels have no standard names. A label such as S1--D1 is assigned per montage eith no spatial criteria and therefore might refer to a different cortical region in each dataset. Recently, several EEG foundation models have used the physical location of the sensors to harmonize the heterogeneity of EEG montages, encoding the 3D coordinates of each electrode in a common space, so that one model can process any montage~\cite{doner2026luna,el2026reve,fang2026neuript,hong2026cortexbridge,broustail2026lumamba}. The same strategy can be applied to fNIRS, where the coordinates are the only spatial reference across probe layouts. GeoNIRS is the first model to use it. It encodes the position of each channel, and removing or shuffling the coordinates reduced its accuracy~\cite{li2026geonirs}. Therefore, we included the 3D position of every channel in every curated dataset, which allows the montage-agnostic approach of EEG foundation models to be extended to fNIRS. 

The third group holds 15 datasets that did not meet the inclusion criteria but provide both HbO and HbR and at least 2D optode coordinates: 13 with a 2D probe layout and 2 with 3D coordinates in an unknown spatial frame. We release their loaders as well, since they can serve models that do not require 3D positions or that can handle 2D and 3D coordinates together (Appendix~\ref{sec:rejected}).

In summary, our contributions are:
\begin{itemize}
  \item \textbf{A large and diverse pretraining corpus} of 61 public fNIRS datasets, 2,577 participants and 1,836~h, more than twice the largest pretraining set used for fNIRS to date (Table~\ref{tab:summary}).
  \item \textbf{Channel positions in a common space.} Every channel of the corpus is given its position in MNI space (Section~\ref{CS}).
  \item \textbf{A downstream benchmark} of six held-out datasets with labeled epochs, covering three clinical groups and three experimental conditions (Section~\ref{sub:downstream}).

\end{itemize}

\section{Methods}
\subsection{Search Strategy}

We conducted a comprehensive search to identify candidate datasets. OpenNeuro~\cite{markiewicz2021openneuro}, Zenodo\footnote{\url{https://zenodo.org}}, figshare\footnote{\url{https://figshare.com}}, the Science Data Bank\footnote{\url{https://www.scidb.cn}}, Mendeley Data\footnote{\url{https://data.mendeley.com}} and Dryad\footnote{\url{https://datadryad.org}} were queried with the terms ``fNIRS'' and ``near-infrared spectroscopy''. fNIRS articles were additionally retrieved from Google Scholar, arXiv, IEEE Xplore and Frontiers with the same terms, and their data-availability statements were followed to the corresponding repositories. In this way, datasets were located on the Open Science Framework~\cite{foster2017open}, Deep Blue Data (University of Michigan)\footnote{\url{https://deepblue.lib.umich.edu/data}}, the Donders Repository\footnote{\url{https://data.ru.nl}}, DR-NTU (Nanyang Technological University)\footnote{\url{https://researchdata.ntu.edu.sg}}, NITRC\footnote{\url{https://www.nitrc.org}}, PhysioNet\footnote{\url{https://physionet.org}}, G-Node GIN\footnote{\url{https://gin.g-node.org}}, Borealis\footnote{\url{https://borealisdata.ca}}, PsychArchives\footnote{\url{https://www.psycharchives.org}}, Hugging Face\footnote{\url{https://huggingface.co}} and the BBCI website of Technische Universit\"at Berlin\footnote{\url{https://doc.ml.tu-berlin.de}}.

\subsection{Selection Criteria}
To ensure that every dataset can serve the development of fNIRS foundation models, datasets were selected according to three criteria: the fNIRS technique, the type of signal shared, and the optode positions.

\paragraph{fNIRS technique.} Only continuous-wave (CW) fNIRS was considered. CW instruments measure only the intensity of the light re-emerging from the tissue, whereas time- and frequency-domain instruments additionally resolve its time of flight at the cost of more complex and expensive instrumentation~\cite{Scholkmann2014}. The vast majority of commercial fNIRS instruments are therefore CW systems, which are comparatively inexpensive, wearable and suited to recordings in everyday settings~\cite{Scholkmann2014}. 

\paragraph{Signals.} The files had to contain the unprocessed time series of each recording session, either raw light intensity or optical density, without any subsequent processing such as filtering or artifact correction. The only exception was made for datasets whose acquisition software, such as that of the Shimadzu LABNIRS instruments, stores HbO and HbR concentrations directly and releases no intensity or optical density. These were accepted when both HbO and HbR were provided in the same unprocessed form. Datasets that released only HbO, or signals that had been processed, were rejected.

\paragraph{Optode positions.} The positions had to be three-dimensional and convertible to MNI space (Section~\ref{CS}). These conditions were considered satisfied when the positions were reported in MNI coordinates, when they had been digitized together with anatomical landmarks, or when each optode was identified with a position of the 10--20 or 10--5 system. Positions available only as a two-dimensional layout, or in an undocumented coordinate frame without landmarks did not satisfy this criterion. The 15 datasets rejected on this ground are listed in Appendix~\ref{sec:rejected}.

\subsection{Coordinate System}\label{CS}
Human brains vary in size, shape and folding, which makes direct comparisons across individuals difficult. The MNI coordinate system addresses this with a standard template, built by registering the MRI scans of several adults. Data from different individuals can then be expressed and compared in this common space.

In fMRI studies, structural images are normalized to the MNI template through linear and nonlinear transformations, and functional data are mapped into the same space~\cite{tsuzuki2007virtual}. In contrast, fNIRS is recorded on the head surface, usually without a structural image. Optode positions are therefore registered to MNI space from head-surface landmarks, either by digitizing them~\cite{singh2005spatial} or by simulating the optode placement on a template head~\cite{tsuzuki2007virtual}. In this work, we used the MNI coordinates reported by the authors when available and otherwise registered the optode positions to MNI space with the procedures described below. Each channel was located at the midpoint between its source and detector on the scalp~\cite{tsuzuki2007virtual}. Some datasets do not provide scalp positions but the cortical projection of each channel (the point on the cortical surface directly below the scalp midpoint). For these datasets we used the cortical coordinates as given.

\begin{figure}[t]
    \centering
    \includegraphics[width=\linewidth]{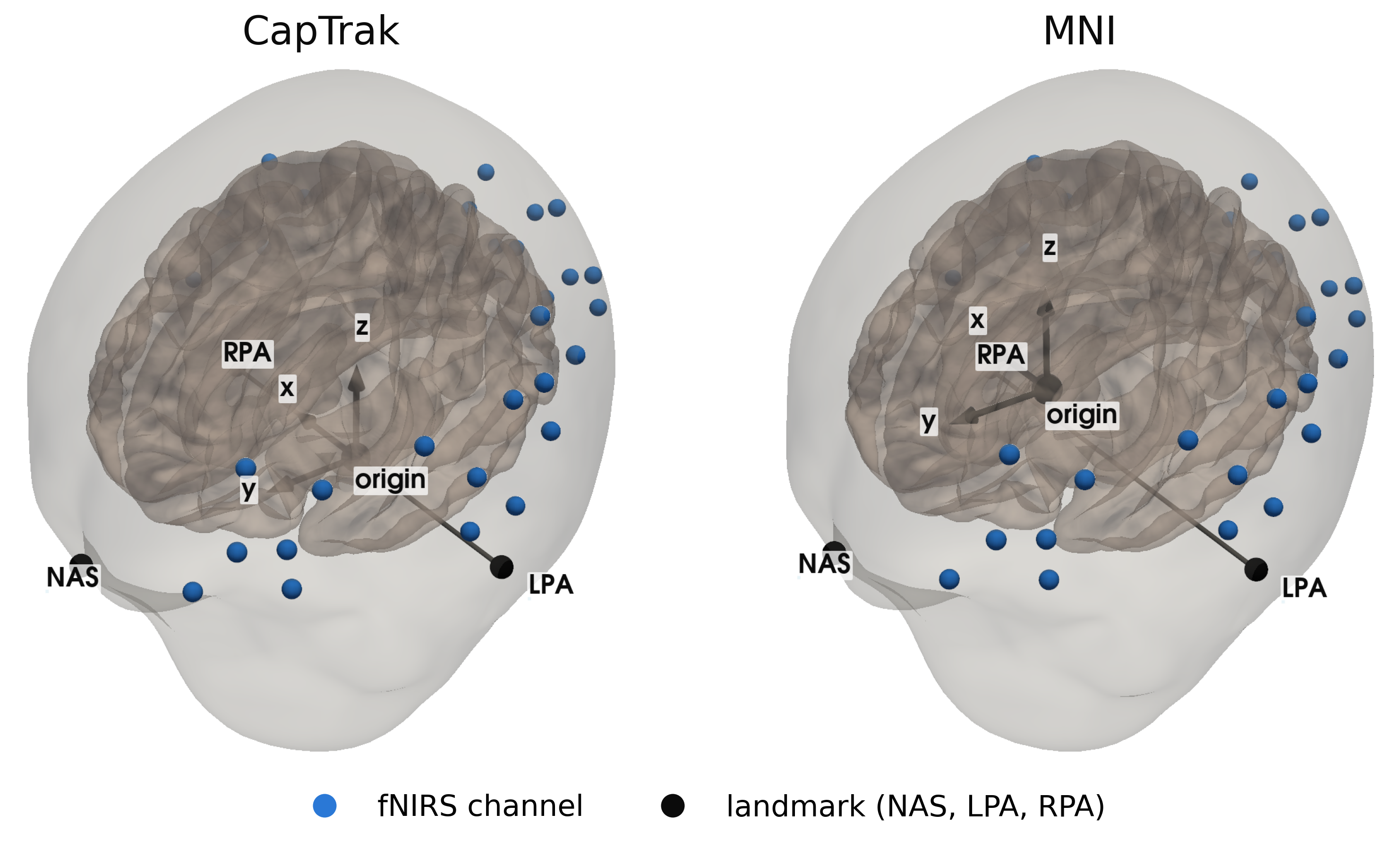}
    \caption{Channel positions of a fNIRS dataset in two coordinate systems. \textbf{Left:} CapTrak is defined by three head landmarks. The $x$-axis runs from the left preauricular point (LPA) through the right preauricular point (RPA), the $y$-axis is orthogonal to it and passes through the nasion (NAS), and the $z$-axis points upwards, so the origin lies approximately between the ears. \textbf{Right:} MNI coordinate system with its origin at the anterior commissure and axes pointing right ($x$), anterior ($y$) and superior ($z$).}
    \label{fig:coordinate_frames}
\end{figure}

Datasets that report positions in CapTrak (Fig.~\ref{fig:coordinate_frames}), the BIDS coordinate system with RAS orientation ($x$ toward the right, $y$ toward the front, $z$ toward the top of the head) and its origin between the left and right preauricular points~\cite{appelhoff2019mne}, were converted to MNI space with \texttt{mne.head\_to\_mni}~\cite{gramfort2013meg}, using the head--MRI transform of \textit{fsaverage}, the FreeSurfer average brain. Datasets that provide optode positions digitized on each participant's head, together with the nasion and the preauricular points, were registered to \textit{fsaverage} with \texttt{mne.coreg.Coregistration}~\cite{gramfort2013meg}, which aligns the template head to the three landmarks and then fits it to the optode positions with a rotation, a translation and a uniform scaling factor. In both cases the resulting positions are in the MRI coordinate system of \textit{fsaverage}, which corresponds to MNI space.

\subsection{Minimal processing}\label{sec:processing}

All recordings were processed with the same minimal pipeline, implemented in MNE-Python~\cite{gramfort2013meg} . The pipeline has four steps: handling of missing sample points, conversion to hemoglobin concentrations, high-pass filtering and resampling. Motion correction, short-channel regression and low-pass filtering were not applied, and no channels were rejected on the basis of signal quality.

\paragraph{Missing samples.} Samples stored as zero, negative or non-finite values were treated as missing. Missing segments at the beginning or end of a recording were removed if they lasted 60~s or less. Otherwise, the affected channel was discarded. Gaps of 1~s or less within a recording were filled by linear interpolation and channels with longer gaps were discarded.

\paragraph{Hemoglobin concentrations.} Raw intensity was converted to optical density as the negative natural logarithm of the intensity normalized by its temporal mean. Concentration changes of HbO and HbR were computed with the modified Beer--Lambert law~\cite{Delpy1988}, using the molar extinction coefficients tabulated in MNE-Python, the source--detector distance of each channel and a differential path-length factor (DPF) of 6.0 at both wavelengths. The DPF is a dimensionless factor that accounts for the increase of the optical path length over the source--detector distance caused by scattering in tissue~\cite{Delpy1988}. In continuous-wave fNIRS it cannot be measured and is usually taken as a constant of about 6~\cite{duncan1995optical}, which we used for all datasets.

\paragraph{Filtering.} Concentration signals were high-pass filtered at 0.01~Hz, the lower cutoff most commonly used in fNIRS studies~\cite{Pinti2019}, to remove instrumental drift and very slow vasomotor fluctuations. The filter is a fourth-order Butterworth filter applied forward and backward in time, which cancels the frequency-dependent phase delay of infinite impulse response filters and preserves the timing of the signals~\cite{Pinti2019,yucel2021best}. No low-pass filter was applied, so that the cardiac ($\sim$1\,Hz), respiratory ($\sim$0.2--0.5\,Hz) and Mayer-wave ($\sim$0.1\,Hz) components of the signal are retained \cite{luke2021characterization}.

\paragraph{Resampling.} The filtered signals were resampled to 7.5~Hz. The original sampling rates ranged from 2.6 to 75~Hz. Of the 67 datasets, 51 were acquired above 7.5~Hz, 12 below (2.6--5.7~Hz) and 4 contain recordings at both lower and higher rates. The rate of 7.5~Hz was therefore selected to minimize the number of datasets requiring upsampling.

\section{Curated Datasets}

The search identified more than 90 candidate datasets, of which 67 met the selection criteria. Together they comprise 1{,}955~h of fNIRS data from 2{,}798 participants, counted separately in each dataset in which they appear. The remaining datasets were excluded because they lacked three-dimensional optode positions (15 datasets, listed in Appendix~\ref{sec:rejected}) or because they provided only HbO or processed signals.

Of the 67 datasets, 61 form the pretraining corpus (1{,}836~h, 2{,}577 participants), summarized by task category in Table~\ref{tab:summary}. The other six are held out for downstream evaluation (119~h, 221 participants; Table~\ref{tab:downstream}). They were selected from the downstream benchmark of GeoNIRS~\cite{li2026geonirs}, an fNIRS foundation model pretrained on up to 22 public datasets and adapted to held-out tasks by linear probing, low-rank adaptation or fine-tuning. Adopting the same downstream datasets allows models pretrained on the curated corpus to be evaluated on the same tasks as GeoNIRS. GeoNIRS evaluates eight downstream tasks; two of its datasets were not included, one because it provides only two-dimensional optode positions~\cite{peng2018morphine} and the other because access is restricted~\cite{raible2026all}.

\subsection{Downstream tasks}\label{sub:downstream}
The downstream datasets were processed with the same pipeline as the pretraining corpus (Section~\ref{sec:processing}). Each recording therefore provides HbO and HbR concentration changes at 7.5~Hz, with channel positions in MNI space. Unlike the pretraining corpus, each recording also carries the onset, duration and label of every experimental condition.

Event onsets were taken from the event files or stimulus markers distributed with each dataset, and each event was labeled according to its classification task. In three tasks the label is the group of the participant: healthy versus intracerebral hemorrhage (ICH)~\cite{shi2025hefmi,shi2025hefmidata}, normal hearing versus cochlear implant (CI) use~\cite{balint2025fnirs,balint2025speechdata}, and healthy versus tinnitus~\cite{satish2026tinnitusdata}. In the other three tasks the label is the condition of the trial. In the movement task~\cite{medvedeva2026multisession,medvedeva2025strokedata}, each trial is labeled as movement or no movement. In the action-mode task~\cite{li2020brain,li2025mirrordata}, trials are labeled as observation or execution of simple and complex hand actions, and the visual control condition is excluded. In the semantic-category task~\cite{rybavr2025simultaneous,rybar2025animalsdata}, trials are labeled as animal or tool according to the concept presented.

One epoch was extracted per labeled event. Each epoch covers the entire condition, as described in the original publication or dataset:
\begin{itemize}
  \item \textbf{ICH}: the 10-s motor imagery period, from 2 to 12~s after the onset of the visual cue~\cite{shi2025hefmi}.
  \item \textbf{CI status}: the 13-s sentence stimulus~\cite{balint2025fnirs}.
  \item \textbf{Tinnitus}: the 12-s broadband noise block~\cite{satish2026tinnitusdata}.
  \item \textbf{Movement}: 12~s from the onset of each flash~\cite{medvedeva2026multisession,medvedeva2025strokedata}.
  \item \textbf{Action mode}: the 4-s action trial~\cite{li2020brain}.
  \item \textbf{Semantic category}: 14.3~s from the onset of each image, covering the image, a mask and the four 3-s mental tasks performed on the object (silent naming and visual, auditory and tactile imagery)~\cite{rybavr2025simultaneous}.
\end{itemize}

\begin{table}[t]
\centering
\footnotesize
\setlength{\tabcolsep}{3pt}
\caption{\textbf{Overview of the pretraining corpus}. The 61 pretraining datasets are grouped into task categories and subcategories according to the experimental paradigm. For each subcategory, the table lists the participant groups, the number of datasets and participants, the total recording time and the publications describing the datasets, or the data records when no publication is available.}
\label{tab:summary}
\begin{tabular}{@{}>{\raggedright\arraybackslash}p{1.6cm}>{\raggedright\arraybackslash}p{2.6cm}>{\raggedright\arraybackslash}p{2.5cm}rrr>{\raggedright\arraybackslash}p{2.3cm}@{}}
\toprule
\textbf{Task category} & \textbf{Subcategory} & \textbf{Population} & \textbf{Datasets} & \textbf{Participants} & \textbf{Hours} & \textbf{Ref.} \\
\midrule
Resting state & Awake rest & Healthy, AD, MCI & 3 & 233 & 90.78 & \cite{yucel2025quantifying,Goodrick2025,Butters2026} \\
 & Sleep, anesthesia or bed rest & Healthy, acute stroke, general anesthesia & 3 & 120 & 41.83 & \cite{blanco2022open,bergonzi2026disrupted,saini2026anesthesiadata} \\
\midrule
Social interaction & Adult interaction & Healthy & 5 & 342 & 440.62 & \cite{moffat2026longitudinal,chen2025fnirs,zheng2026coopvmdata,hu2025dataset,hakim2026investigating} \\
 & Family interaction & Healthy & 2 & 172 & 45.93 & \cite{bizzego2022dataset,azhari2020physical} \\
\midrule
Emotion & Emotion induction & Healthy & 2 & 77 & 40.51 & \cite{ge2024empathy,NeurIPS2025_REFED} \\
 & Emotion regulation & Healthy & 1 & 25 & 17.95 & \cite{caifengdifferential} \\
\midrule
Speech & Auditory perception & Healthy & 4 & 95 & 77.52 & \cite{duwadi2026decoding,Steinmetzger2022,luke2022human,luke2021analysis} \\
 & Language processing & Healthy, CI users & 6 & 516 & 198.51 & \cite{pasquinelli2023development,sun2022morphological,Yu2025,Elliott2026,Shader2021,Mutlu2023} \\
 & Language production & Healthy & 1 & 33 & 34.59 & \cite{Schroeder2023} \\
\midrule
Motor & Motor execution & Healthy & 5 & 80 & 49.57 & \cite{Cockx2023,carlton2026surface,khan2026open,mesquita2025freshmotor,luke2021tappingdata} \\
 & Motor imagery and observation & Healthy, stroke & 5 & 105 & 155.88 & \cite{isaev2024multiple,Shin2017,yi2025multi,feng2026multimodal,almulla2022observation} \\
 & Gait & Healthy, PD & 2 & 64 & 23.69 & \cite{Cockx2024,minhas2024enhancing} \\
 & Surgical skills & Healthy & 2 & 90 & 159.99 & \cite{kamat2023assessment,torrescuevas2025resident} \\
 & Exercise and pain stimulation & Healthy & 1 & 42 & 27.67 & \cite{geisler2026profound,geisler2025athletedata} \\
\midrule
Cognitive & Attention and executive control & Healthy, brain injury & 4 & 184 & 144.71 & \cite{zhang2025visuomotor,Ding2025,zhu2025multitaskdata,anderson2025high} \\
 & Mental workload & Healthy & 4 & 99 & 125.77 & \cite{zhang2024fnirs,ajra2026open,Hamann2022,hamann2026examining} \\
 & Working memory & Healthy & 3 & 64 & 64.39 & \cite{shin2018simultaneous,Sztuka2025,Csipo2021} \\
 & Judgment and decision-making & Healthy & 3 & 144 & 28.94 & \cite{Cheng2026,KawabataDuncan2019,ding2026wired} \\
\midrule
Visual & Movies and localizers$^{a}$ & Healthy & 3 & 19 & 53.78 & \cite{tripathy2024mapping,fogarty2025functional,Fehner2026} \\
 & Visual stimulation & Healthy, AD, MCI & 2 & 73 & 13.25 & \cite{Butters2026,rocco2026comparison} \\
\midrule
Total & & & 61 & 2{,}577 & 1{,}835.88 & \\
\bottomrule
\end{tabular}
\par\smallskip
\parbox{\linewidth}{\footnotesize $^{a}$Functional localizers are short tasks that activate a specific cortical area and are used to identify the location of this area in each participant.}
\end{table}

\begin{table}[t]
\centering
\footnotesize
\setlength{\tabcolsep}{3pt}
\caption{\textbf{Downstream tasks}. The six datasets are held out from the pretraining corpus.}
\label{tab:downstream}
\begin{tabular}{@{}>{\raggedright\arraybackslash}p{2.2cm}>{\raggedright\arraybackslash}p{3.6cm}>{\raggedright\arraybackslash}p{3.6cm}rr@{}}
\toprule
\textbf{Task} & \textbf{Classes} & \textbf{Population} & \textbf{Participants} & \textbf{Hours} \\
\midrule
ICH~\cite{shi2025hefmi} & Healthy vs.\ ICH patients & ICH, healthy controls & 37 & 25.61 \\
CI status~\cite{balint2025fnirs} & Normal hearing vs.\ CI & CI users, normal-hearing controls & 72 & 32.00 \\
Tinnitus~\cite{satish2026tinnitusdata} & Control vs.\ tinnitus  & Tinnitus, healthy controls & 63 & 17.58 \\
Movement~\cite{medvedeva2026multisession} & Movement vs.\ no movement & Stroke & 16 & 23.20 \\
Action mode~\cite{li2020brain} & Observation vs.\ execution & Healthy & 21 & 5.86 \\
Semantic category~\cite{rybavr2025simultaneous} & Animal vs.\ tool & Healthy & 12 & 14.71 \\
\midrule
Total & & & 221 & 118.96 \\
\bottomrule
\end{tabular}
\end{table}

\section{Discussion}

\subsection{Rationale for the selection criteria}

\paragraph{HbO and HbR.}
fNIRS measures both HbO and HbR, whereas the BOLD signal of fMRI reflects mainly HbR~\cite{pinti2020present}. Through neurovascular coupling, an active region receives more oxygen than it consumes, so HbO increases and HbR decreases~\cite{pinti2020present}. Therefore, the two signals are usually negatively correlated, while head motion makes them more positively correlated~\cite{cui2010functional}. HbO has the larger amplitude and the higher signal-to-noise ratio but is more affected by systemic physiology. In contrast, HbR has less statistical power but is more robust to systemic changes~\cite{pinti2020present}.

We therefore excluded datasets that provide only HbO. Although HbO exhibits the larger task-evoked amplitude and yielded higher accuracy than HbR when used alone in a recent ablation study~\cite{wang2026domain}, the literature does not establish a consistently superior chromophore for decoding tasks. For speech decoding, the two signals yielded statistically indistinguishable accuracy~\cite{kamavuako2018classification}; for mental arithmetic, their performance differed by less than two percentage points~\cite{han2022functional}; and for finger tapping and motor imagery, the more informative chromophore varied with the task and the type of hand-crafted features~\cite{nazeer2020enhancing,qiu2022improved}. Moreover, combining both signals improved classification accuracy over either signal alone~\cite{nazeer2020enhancing,qiu2022improved}. Since the relative value of HbO and HbR depends on the decoding task, retaining both is the appropriate choice for a corpus intended to support a broad range of downstream applications.

\paragraph{Channel positions in MNI space.}

A model pretrained on heterogeneous datasets must account for the spatial origin of each signal, because sensor placement differs between datasets. EEG foundation models have addressed this problem in two ways. Early models such as BIOT and LaBraM maintain a lookup table containing one learned embedding per channel name~\cite{Yang2023,jiang2024large}. Spatial information is available to the model only through these embeddings, so the model can represent only channels whose names were present during pretraining. A channel with an unseen name, or a dataset that follows a different naming convention, has no entry in the table and cannot be used without additional training~\cite{el2026reve}. More recent models instead derive the encoding of each channel from its three-dimensional coordinates~\cite{doner2026luna,el2026reve,fang2026neuript,broustail2026lumamba}. Because every sensor has coordinates, these models accommodate arbitrary layouts, including those absent from the pretraining data. In NeurIPT, coordinate-based encoding outperformed learned channel embeddings on every dataset evaluated~\cite{fang2026neuript}.

A name-based table is unsuitable for fNIRS. A channel is defined by a source and a detector, and its label, such as S1--D1, is assigned within each montage, so identical labels correspond to different cortical regions across datasets. GeoNIRS encodes the position of each channel, and removing or permuting these positions reduces its accuracy, which indicates that the model exploits the recording location of each signal~\cite{li2026geonirs}. In our corpus, every channel is assigned a three-dimensional position in MNI space. MNI space is a standardized anatomical coordinate frame defined on a template brain, so a given coordinate denotes the same cortical location regardless of the subject, device or laboratory from which the recording originates. Datasets release positions in a variety of frames, including digitizer coordinates relative to the subject's own head, device-specific template frames and 2D probe layouts. Coordinates expressed in different frames are not comparable. The same numerical position may refer to different cortical regions, and the same region may receive different positions, so a model could not learn a consistent relation between position and signal across datasets. Expressing all positions in MNI space allows a model to learn both the geometry of each probe and its placement on the head. The 15 datasets without such positions, comprising 182~h of recordings, are released separately (Appendix~\ref{sec:rejected}).

\subsection{Minimal processing}

Apart from the handling of missing samples, the pipeline converts raw intensity to optical density and then to HbO and HbR with the modified Beer--Lambert law, applies a 0.01~Hz high-pass filter and resamples to 7.5~Hz. The two conversions are standard in continuous-wave fNIRS analysis. Raw intensity has no common scale across devices, since its level depends on the source power, the detector gain and the optical coupling of each optode to the scalp. Optical density removes these multiplicative factors by expressing each channel relative to its own mean. The modified Beer--Lambert law then converts the optical density changes measured at two wavelengths into concentration changes of HbO and HbR, accounting for the source--detector distance and the differential pathlength factor~\cite{Delpy1988,pinti2020present}.

The high-pass filter removes slow instrumental drift. Its 0.01~Hz cutoff is the most common in fNIRS studies~\cite{Pinti2019}, and a higher cutoff risks attenuating the hemodynamic response itself~\cite{yucel2021best}. Most studies also apply a low-pass filter, with cutoffs between 0.09 and 0.5~Hz~\cite{Pinti2019}, to suppress systemic oscillations such as the cardiac pulse and respiration~\cite{Scholkmann2014}. We omitted this step because these oscillations are not purely noise: their amplitude is modulated by the task, the slower components vary with posture and age~\cite{Scholkmann2014}, and the cardiac pulse indicates how well each optode is coupled to the scalp~\cite{yucel2021best}. Moreover, the extent to which physiological components affect pretraining and downstream performance has not been established.

\subsection{Limitations and Future work}

Although every channel position is in MNI space, the positions were obtained in three ways. In 14 datasets the optodes were digitized on each participant’s head or placed on a participant-specific MRI head model, and we registered them to MNI space. In 15 datasets the authors provide MNI coordinates obtained by registering a standard layout; in some of these, the coordinates are the cortical projection of each channel, 18–30 mm below the scalp, rather than its scalp position. In the remaining datasets the optodes are identified only by their 10–20 or 10–5 location, and we assigned each one the template coordinates of that location, identical for all participants. Coverage is also uneven. Most sparse montages cover the prefrontal cortex, and whole-head coverage is provided by only a few high-density datasets.

As future work, we intend to integrate this corpus into a shared Neuro-AI framework such as NeuralSet~\cite{king2026neuralset}, which supports fNIRS. The corpus fixes the data and their preprocessing, but windowing, the partition of participants into training and test sets, and the evaluation protocol of the downstream tasks are still left to each study. Implementing these steps once, within a common framework, would hold them fixed across studies, so that differences in performance can be attributed to the architecture and pretraining strategy rather than to the experimental setup.

\section{Conclusion}

In this study, we reviewed more than 90 public fNIRS datasets and assembled a pretraining corpus from 61 of them, comprising 2,577 participants and 1,836~h of recordings from several acquisition systems, seven task categories and participants ranging from infants to older adults. Every recording provides HbO and HbR concentration changes at 7.5~Hz after minimal processing, and every channel has a three-dimensional position in MNI space. Six held-out datasets with labeled epochs form a downstream benchmark, and loaders for 15 further datasets without 3D positions are released separately. All loaders rebuild each dataset from its public repository and apply the same processing, so the corpus can be reproduced from the code and extended as new datasets appear. We expect the corpus to lower the barrier to developing fNIRS foundation models.

\bibliographystyle{plainnat}
\bibliography{ref}

\clearpage
\appendix

\section{Pretraining corpus}\label{sec:datasets}

The 61 datasets of the pretraining corpus are described below, grouped by the task categories of Table~\ref{tab:summary}. The six downstream datasets are described in Appendix~\ref{sec:downstream_datasets}.

\subsection{Resting state}\label{app:resting}

\paragraph{Hair and skin characteristics.}
This dataset contains recordings from 115 adults recruited to cover a wide range of hair and skin characteristics~\cite{yucel2025quantifying}. Each participant completed two 3-min resting-state runs, followed by a 6-min block of left- and right-hand ball squeezing. Data were acquired with two CW NIRSport2 devices (NIRx, Berlin, Germany) recording simultaneously, one per hemisphere, each with 8 LED sources (760/850~nm) and 16 silicon photodiode detectors forming 18 long channels (15--34~mm) and 8 short channels (about 8~mm) over the forehead, the side and the back of the head, sampled at 10.2~Hz. Because the two devices were not synchronized, we treated the recording of each hemisphere as a separate recording. The files provide the NIRx ICBM152 template positions of the cap, which we used as given. The dataset is available on OpenNeuro~\cite{yucel2025inclusiondata}.

\paragraph{Yoga intervention.}
This dataset contains recordings from adults in a 30-day yoga intervention group and a control group, acquired before and after the 30-day period during 6~min of rest, 9~min of affective image viewing and 4.5~min of breath holding~\cite{Goodrick2025}. Data were acquired with a CW device (not reported) with 10 sources (757/843~nm) and 8 detectors forming 24 channels over the prefrontal cortex, sampled at 25 or 75~Hz. The two short-separation channels, which have no coordinates, were excluded, leaving 22 channels per recording, and five recordings without a participant identifier were not used, resulting in 232 recordings from 56 participants. The authors provide MNI coordinates for each channel, which we used as given; the source--detector separation was set to 30~mm. The dataset is publicly available on Zenodo~\cite{goodrick2026yogadata}.

\paragraph{Resting state in Alzheimer's disease and mild cognitive impairment.}
This dataset comprises 5-min eyes-closed resting-state recordings from 65 older adults: 21 with Alzheimer's-type dementia, 22 with Alzheimer's-type mild cognitive impairment and 22 age-matched healthy controls~\cite{Butters2026}. Data were acquired with a CW LUMO high-density diffuse optical tomography device (Gowerlabs, London, UK) with 12 modules over the bilateral frontal cortex, each with 3 dual-wavelength LED sources (735/850~nm) and 4 photodiode detectors (36 sources and 48 detectors in total), sampled at 12.5~Hz. The files contain all 1728 source--detector pairs, with source--detector separations (SDS) ranging from 10 to over 100~mm. Increasing the SDS increases the sensitivity to cortical tissue but reduces the detected light intensity and, consequently, the signal-to-noise ratio~\cite{wang2019investigation}. SDS of 25--35~mm have therefore been recommended for the adult head as a tradeoff between both effects~\cite{wang2019investigation}. We retained the pairs with SDS $\leq$ 35~mm, a criterion that we applied to all high-density datasets, yielding up to 396 channels per recording, of which approximately 108 are short-separation channels (SDS $<$ 12~mm). Optode positions were digitized for each participant by photogrammetry, together with the nasion, inion, preauricular points and vertex. The optode positions and the nasion and preauricular points were registered to MNI space with \texttt{mne.coreg.Coregistration} (Section~\ref{CS}). Three participants with inconsistent landmark coordinates were excluded, resulting in 62 recordings. The dataset is publicly available on OpenNeuro~\cite{butters2026cambridgedata}.

\paragraph{Infant resting state.}
This dataset contains resting-state recordings from 104 healthy, full-term 4-month-old infants, acquired during natural sleep while the infants leaned on their parents' lap~\cite{blanco2022open}. Data were acquired with a CW NIRScout device (NIRx, Berlin, Germany) with 52 channels (16 sources, 760/850~nm, and 24 detectors; source--detector separations of 20--45~mm) over frontal, temporal, parietal and occipital regions of both hemispheres, sampled at 8.93~Hz. The optodes were placed according to the international 10--20 system, using the nasion, inion and preauricular points as landmarks. Following the original analysis, we excluded the 5 infants that the authors discarded because of insufficient data quality, keeping 99 recordings. We also discarded the 6 occipital channels, which were prone to artifacts because the back of the infants' heads rested against the parent's body, leaving 46 channels. No 3D optode positions were recorded; the dataset provides only the 10--20 position of each source and detector, so we assigned each optode the MNI coordinates of its 10--20 position on the NIRx ICBM152 template. The dataset is available on the Open Science Framework~\cite{blanco2022osf}.

\paragraph{Bedside resting state after acute ischemic stroke.}
This dataset contains resting-state recordings from 13 patients with acute ischemic stroke in the middle cerebral artery territory, acquired at the bedside in the intensive care unit within 72~h of stroke onset, resulting in 54 runs of 3--11~min~\cite{bergonzi2026disrupted}. Data were acquired with a custom-built CW HD-DOT system (Washington University in St.~Louis, MO, USA) with 48 LED sources (750/850~nm) and 34 avalanche photodiode detectors arranged in two arrays over both hemispheres, with first- to fourth-nearest-neighbor SDS of 13, 30, 39 and 47~mm, sampled at 10.0~Hz. We retained the pairs with SDS $\leq$ 35~mm, yielding 294 channels per recording. Optode positions are provided on the MNI152 atlas head, adjusted for each participant's cap position and head size, together with the nasion, inion and preauricular points, and were registered to MNI space. The dataset is publicly available on OpenNeuro~\cite{bergonzi2026strokedata}.

\paragraph{General anesthesia.}
This dataset contains simultaneous EEG and fNIRS recordings from 8 patients undergoing general anesthesia, one of whom was recorded in two sessions~\cite{saini2026anesthesiadata}. Each recording spans the induction of anesthesia and the subsequent recovery, and includes a baseline period, verbal, motor and tetanic stimuli, and the loss and return of consciousness. No associated publication was available. The 9 recordings of 20--81~min were acquired with a CW NIRx device (760/850~nm; model not reported) sampled at 7.81~Hz, with three montages of 15, 19 and 20 channels. Optode positions are provided as template positions in the CapTrak coordinate system and were transformed to MNI space with \texttt{mne.head\_to\_mni} (Section~\ref{CS}). The dataset is publicly available on OpenNeuro~\cite{saini2026anesthesiadata}.

\subsection{Social interaction}\label{app:social}

\paragraph{Intergenerational collaborative drawing.}
This dataset contains fNIRS hyperscanning recordings from 122 adults in 61 dyads who met as strangers and attended six weekly sessions of a creative drawing program~\cite{moffat2026longitudinal}: 31 intergenerational dyads, each formed by an older adult and a younger adult, and 30 dyads of two younger adults. In each session, the two members of a dyad drew with oil pastels for 5~min alone and twice for 5~min together on a single sheet of paper, without talking, and then completed a short collaborative task; each member was recorded in a single recording of about 20~min. The data were acquired with CW Photon Cap devices (Cortivision, Lublin, Poland) with 16 LED sources (760/850~nm), 4 of them dedicated to short-separation channels, and 10 silicon photodiode detectors at 10--5 positions over the inferior frontal gyrus and the temporoparietal junction of both hemispheres, forming 22 long channels (20--40~mm) and 4 short channels (8--14~mm), sampled at about 5~Hz. The authors digitized the optode positions in each session from a 3D scan of the participant's head and provided the MNI coordinates of each channel for part of the sessions; we used these coordinates where available and, for the remaining recordings, including all those of the older adults, the template coordinates of the 10--5 positions stored in the files. The authors flagged the channels without cardiac oscillations or with major motion artifacts, identified by visual inspection; we discarded these channels. We also excluded two recordings acquired with an earlier montage that the files do not describe. The dataset is available on OpenNeuro~\cite{moffat2026intergendata}.

\paragraph{Social touch.}
This dataset contains hyperscanning recordings from 47 female dyads, 24 pairs of close friends and 23 pairs of strangers, who sat side by side and viewed negative and neutral pictures, rating each one for valence and arousal~\cite{chen2025fnirs}. A no-touch session was followed by a session in which the partners held hands; each session comprised eight blocks of ten pictures, and resting-state data were recorded at the start and at the end of the experiment. Each participant was recorded with a CW NIRScout device (NIRx, Berlin, Germany) with 20 channels (8 sources, 785/830~nm, and 7 detectors) over the prefrontal cortex, sampled at 7.81~Hz, using the same montage as the parent--child dataset described below. Every optode lies on a 10--20 position, and we assigned it the MNI coordinates of that position on the NIRx ICBM152 template. The dataset is available on the Open Science Framework~\cite{chen2025socialdata}.

\paragraph{Cooperative visuomotor control in dyads.}
This dataset contains fNIRS hyperscanning recordings from 64 healthy young adults in 32 same-sex dyads who colored a picture together using two haptic-feedback devices, in four runs of 10--13~min combining two collaboration modes and two difficulty levels~\cite{zheng2026coopvmdata}; the accompanying publication was not yet available. Data were acquired with a CW NirScan device (Danyang Huichuang Medical Equipment, Danyang, China) at 730, 808 and 850~nm, sampled at 11~Hz, with 26 channels per participant (SDS of 30~mm) over the left and right prefrontal cortex, the left sensorimotor cortex and the right temporoparietal junction. Each recording contains both members of a dyad and was split into one recording per participant, resulting in 256 recordings. The authors provide MNI coordinates for each channel, which we used as given. The dataset is publicly available on the Science Data Bank~\cite{zheng2026coopvmdata}.

\paragraph{Natural conversations about appearance.}
This dataset contains recordings from 31 female participants during three natural conversations: a fat talk about their bodies, preceded and followed by appearance-unrelated topics such as travel and home decoration~\cite{hu2025dataset}. Data were acquired with a CW NIRSport device (NIRx, Berlin, Germany) with 40 channels (16 sources, 760/850~nm, and 16 silicon photodiode detectors; 30-mm source--detector separation) arranged in two $4\times4$ probe boards over frontal and parietal areas, sampled at 7.6~Hz. MNI coordinates are provided for each channel. The dataset is available on the Open Science Framework~\cite{hu2025appearance_fnirs_osf}.

\paragraph{Online face gaze.}
This dataset contains fNIRS recordings from 40 adults forming 20 dyads, with one partner at Yale University and the other at University College London (UCL), who gazed at each other's face over a videoconference, either via a live video stream or a static photograph, in two 2-min conditions repeated twice~\cite{hakim2026investigating}. Data were acquired at both sites with CW LABNIRS devices (Shimadzu, Kyoto, Japan) operating at 780, 805 and 830~nm: at UCL, 88 channels (76 for three participants) over temporal, parietal and occipital regions, with a 30-mm source--detector separation, sampled at 12.34~Hz; at Yale, 128 whole-head channels, with a separation of 27.5 or 30~mm depending on head size, sampled at 8.13~Hz. The released data do not include raw light intensities, only hemoglobin changes computed by the Shimadzu software and expressed as concentration multiplied by optical path length (mM$\cdot$cm). We converted them to concentrations using the path length adopted for the other datasets (30-mm source--detector separation and a differential pathlength factor of 6). Channel positions are provided in MNI coordinates for each participant. For 9 of the 20 Yale participants, the positions were missing or did not match the recorded channels; we excluded these participants, retaining 20 UCL and 11 Yale participants. The dataset is available on Dryad~\cite{hirsch2026hyperscanningdata}.

\paragraph{Parent--child interaction.}
This dataset contains simultaneous recordings from 62 parent--child dyads, with children aged 34--55 months~\cite{bizzego2022dataset}. Each dyad completed two sessions: in the first, the child sat on the parent's lap while both watched three 1-min animated clips of different emotional valence; in the second, parent and child played freely with toys for 10~min. Each participant was recorded with a CW NIRSport device (NIRx, Berlin, Germany) with 20 channels (8 sources, 760/850~nm, and 7 silicon photodiode detectors; 30-mm source--detector separation) over the prefrontal cortex, sampled at 7.81~Hz, with the same montage for adults and children. The files provide the NIRx template positions of the montage, in which every optode lies on a 10--20 position; we assigned each optode the MNI coordinates of its position on the NIRx ICBM152 template. In 76 of the 248 recordings, samples were missing and were handled with the procedure of Section~\ref{sec:processing}. The dataset is available from the data repository of Nanyang Technological University~\cite{gabrieli2022parentdata}.

\paragraph{Co-parenting couples.}
This dataset contains recordings from 24 mother--father couples with a child aged 4 years or younger, who listened to infant and adult vocalizations and to static noise, each presented three times for 15~s~\cite{azhari2020physical}. Each couple performed the task twice: together in the same room, and separately in different rooms at different times. Each parent was recorded with a CW NIRSport device (NIRx, Berlin, Germany) with 20 channels (8 sources, 760/850~nm, and 7 silicon photodiode detectors; 30-mm source--detector separation), using the same prefrontal montage as the parent--child dataset, sampled at 7.81~Hz; the optode positions were assigned in the same way. In 42 of the 96 recordings, samples were missing and were handled in the same way. The dataset is available from the data repository of Nanyang Technological University~\cite{esposito2019spousedata}.

\subsection{Emotion}\label{app:emotion}

\paragraph{Empathy and emotional responses to Pipa music.}
This dataset contains recordings from university students with high or low trait empathy, selected with the Interpersonal Reactivity Index, who listened to Chinese Pipa music~\cite{ge2024empathy}. Participants listened to 14 Pipa excerpts, 7 with positive and 7 with negative emotional valence. After each excerpt, they rated its valence, arousal and tension, as well as 6 basic and 9 aesthetic emotions. Data were acquired with a CW NIRScoutX device (NIRx, Berlin, Germany) with 24 LED sources (760/850~nm) and 24 detectors at standard 10--5 positions, forming 81 channels with SDS of 20--44~mm over the prefrontal, temporal and parietal cortex. The recordings were sampled at 2.60~Hz, the rate obtained with 24 time-multiplexed sources. All channels were retained. Optode positions are the NIRx ICBM152 template positions of the probe stored in the files in CapTrak coordinates, identical for all participants; the digitized positions reported in the paper are not included in the dataset. They were transformed to MNI space with \texttt{mne.head\_to\_mni} (Section~\ref{CS}). The dataset is publicly available on the Science Data Bank~\cite{ge2024pipadata}.

\paragraph{Emotional film clips.}
This dataset contains recordings from 32 adults who watched 15 emotional film clips of 1--3~min~\cite{NeurIPS2025_REFED}. EEG and fNIRS were recorded simultaneously; the fNIRS data were acquired with a CW LABNIRS device (Shimadzu, Kyoto, Japan) with 51 channels (18 sources, 780/805/830~nm, and 18 detectors; 30-mm source--detector separation) over the prefrontal cortex and the left and right temporal lobes, sampled at 47.62~Hz. As for the online face-gaze dataset (Appendix~\ref{app:social}), we used the hemoglobin changes computed by the Shimadzu software, expressed as concentration multiplied by optical path length, and converted them to concentrations with a 30-mm separation and a differential pathlength factor of 6. The authors report the 3D channel coordinates in a head-centered frame with the $x$-axis toward the left ear and the $y$-axis toward the back of the head; we reoriented them to the CapTrak convention and converted them to MNI space as described in Section~\ref{CS}. For each participant, we discarded the channels that the authors marked as bad, keeping 35--49 of the 51 channels. Each clip was treated as a separate recording, for a total of 480 recordings. The dataset is available on Hugging Face~\cite{ning2025refeddata}.

\paragraph{Detached and positive cognitive reappraisal.}
This dataset contains recordings from 25 adults who performed an emotion-regulation task comparing two cognitive reappraisal strategies, detached and positive reappraisal~\cite{caifengdifferential}. Each participant was recorded once for 40--45~min, except one participant whose first session was stopped after 4.6~min and repeated; both sessions were retained, resulting in 26 recordings. Data were acquired with a CW NIRScoutX device (NIRx, Berlin, Germany) with 16 laser sources (785/830~nm) and 12 avalanche photodiode detectors at standard 10--20 positions, forming 41 channels with SDS of 30~mm over the prefrontal and the anterior temporal lobes, sampled at 3.91~Hz. The optode 10--20 labels were obtained from the probe file, and each optode was assigned the MNI position of its label on the NIRx ICBM152 template, identical for all participants. The dataset is publicly available on the Science Data Bank~\cite{wang2023reappraisaldata}.

\subsection{Speech}\label{app:speech}

\paragraph{Audiovisual spatial attention in a cocktail-party paradigm.}
This dataset contains recordings from 38 adults who performed spatial-attention tasks in a cocktail-party paradigm~\cite{duwadi2026decoding}. Data were acquired with a CW, wearable, whole-head high-density NINJANIRS22 system (Boston University, Boston, MA, USA). The system has 56 sources and 144 detectors (760/850~nm), forms 567 source--detector pairs with SDS of 16--42~mm, and samples at 8.99~Hz. We kept the 510 pairs with SDS $\leq$ 35~mm. We excluded one recording because its file has no probe description, which left 198 recordings. The optode positions are template positions and the probe layout was designed in AtlasViewer on the Colin27 atlas head. To move the positions to MNI space, we matched 296 10--5 landmarks of the Colin27 head by label to the same landmarks on the ICBM152 template. With MNE-Python (\texttt{mne.coreg.fit\_matched\_points}), we then estimated the rotation, translation and per-axis scaling that best map the Colin27 landmarks onto the ICBM152 ones. We applied this transformation to all optodes. The dataset is publicly available on OpenNeuro~\cite{duwadi2026cocktaildata}.

\paragraph{Voice pitch changes.}
This dataset contains simultaneous fNIRS and EEG recordings from 20 normal-hearing adults who listened passively to 16-s blocks of synthetic German vowels whose pitch contours were either the same for all vowels in a block or varied between vowels, with one session of 61--71~min per participant~\cite{Steinmetzger2022}. Data were acquired with a CW NIRScout device (NIRx, Berlin, Germany) with 8 sources (760/850~nm) and 8 detectors per hemisphere over the auditory cortices, forming 44 channels, of which 40 have an SDS of approximately 30~mm and 4 of approximately 15~mm, sampled at 7.81~Hz (3.91~Hz for one participant). All channels were retained. The optode positions, together with the nasion and preauricular points, were digitized for each participant and registered to MNI space with \texttt{mne.coreg.Coregistration} (Section~\ref{CS}); the Beer--Lambert conversion used the digitized source--detector distances. The dataset is publicly available on the Open Science Framework~\cite{steinmetzger2022voicepitchdata}.

\paragraph{Interaural coherence of sounds.}
This dataset contains recordings from 20 adults who listened passively to 5-s band-pass noises with different degrees of interaural coherence, and to silent control trials, while watching a silent subtitled film~\cite{luke2022human}. Each condition was presented 25 times in random order, and each participant was recorded once for 46--68~min. Data were acquired with a CW NIRScout device (NIRx, Berlin, Germany) with 16 sources (760/850~nm) and 16 detectors forming 88 channels with SDS of 24--65~mm over the superior temporal gyrus and the inferior parietal lobe of both hemispheres, sampled at 7.81~Hz for 10 participants and at 3.91~Hz for the other 10. Optode positions are a template stored in the CapTrak coordinate system, identical for all participants. They were transformed to MNI space with \texttt{mne.head\_to\_mni} (Section~\ref{CS}). The dataset is publicly available on the Open Science Framework~\cite{luke2022interauraldata}.

\paragraph{Passive listening.}
This dataset contains recordings from 17 adults who listened passively to speech, noise and silence in a block design~\cite{luke2021analysis}. Data were acquired with a CW NIRScoutX device (NIRx, Berlin, Germany) with 12 sources (760/850~nm) and 12 avalanche photodiode detectors, plus 8 short-separation detectors at 8~mm, forming 25 long and 8 short channels over the left inferior frontal gyrus, the left and right superior temporal gyri and the occipital lobe, sampled at 5.2~Hz. The optode positions are the NIRx ICBM152 template positions stored in CapTrak coordinates; we converted them back to MNI space as described in Section~\ref{CS}. The dataset is available on OpenNeuro~\cite{luke2025freshaudio}.

\paragraph{Sentence-level prosodic processing in children and adults.}

This dataset contains recordings from 31 adults, 39 children with normal hearing and 6 children with cochlear implants~\cite{pasquinelli2023development}. Participants performed up to three auditory tasks. In the intonation task, they heard pairs of sentences and judged whether the two sentences contained the same words. In the rhyming task, they judged whether two words rhymed. In the oddball task, they listened to single phonemes and pressed a button when they heard an oddball phoneme. This gives 197 recordings of 3--20~min. Data were acquired with a CW6 system (TechEn, Milford, MA, USA) at 690/830~nm and 50~Hz. The cap had six sources and 12 detectors on each hemisphere, giving 46 channels over the bilateral prefrontal, temporal and parietal regions. The dataset includes a file of 46 MNI coordinates without labels. The dataset description states that these are the coordinates of the sources and detectors. However, the cap has 36 optodes and 46 channels, so we treated them as channel coordinates. The paper numbers channels 1--23 on the left hemisphere and 24--46 on the right, as do the MNI coordinates. Channels switched off in the files were removed. The dataset is publicly available on Deep Blue Data~\cite{pasquinelli2021prosodydata}.

\paragraph{Morphological and phonological processing in monolingual and bilingual children.}
This dataset contains recordings from 341 children: English monolinguals and Chinese--English and Spanish--English bilinguals~\cite{sun2022morphological}. The children performed morphological and phonological awareness tasks of 7.2~min in each of their languages. Data were acquired with the same CW6 system (TechEn, Milford, MA, USA) and the same 46-channel cap as the prosody dataset described above, at 690/830~nm and 50~Hz; 68 recordings were sampled at 25~Hz. Most files contain 8 additional source--detector pairs, which have no MNI coordinates and were removed. The 46 remaining channels have the same pairs and the same positions in the 2D drawing as in the prosody dataset, so we used the MNI coordinates of that dataset. Channels switched off in the files were removed. We excluded one truncated file and two files whose time vectors do not match their data. The dataset is publicly available on Deep Blue Data~\cite{hernandez2025morphologicaldata}.

\paragraph{Naturalistic story listening in children.}
This dataset contains recordings from 51 children who listened to the first chapter of \textit{Alice's Adventures in Wonderland}~\cite{Yu2025}. Data were acquired with the same CW6 system (TechEn, Milford, MA, USA) and the same 46-channel cap as the prosody dataset described above, at 690/830~nm and 50~Hz. The files contain 8 additional source--detector pairs, which have no MNI coordinates and were removed. The 46 remaining channels have the same pairs and the same positions in the 2D drawing as in the prosody dataset, so we used the MNI coordinates of that dataset. Channels switched off in the files were removed. The dataset is publicly available on Deep Blue Data~\cite{yu2022alicedata}.

\paragraph{Test--retest lexical decision.}
This dataset contains recordings from 30 English-speaking adults who performed a lexical decision task in two identical sessions one week apart~\cite{Elliott2026}. Participants judged whether real words, pseudowords, consonant strings and false-font strings were real words, interleaved with an active baseline task. Data were acquired with a CW NIRSport2 device (NIRx, Berlin, Germany) with 16 LED sources (760/850~nm) and 16 detectors plus 8 short-separation detectors, forming 42 long channels with SDS of 29--42~mm and 8 short-separation channels with SDS of 7--9~mm over the bilateral inferior frontal, temporal and inferior parietal cortices. The files are sampled at 5.09~Hz. Following the dataset terms of use, two participants who did not consent to secondary use were excluded. We also removed the 15 runs that the authors excluded for having more than eight bad channels, which removed all runs of one participant. This resulted in 264 recordings from 27 participants. Optode positions are the NIRx template positions stored in the files in CapTrak coordinates, identical for all participants. They were transformed to MNI space with \texttt{mne.head\_to\_mni} (Section~\ref{CS}). The dataset is available on Borealis under a CC BY-NC-SA 4.0 license~\cite{elliott2026lexicaldata}.

\paragraph{Auditory-only and visual-only speech.}
This dataset contains recordings from 8 normal-hearing adults~\cite{Shader2021}. Each participant completed 46 trials: 36 story segments (10--16~s each) presented in random order---18 auditory-only trials and 18 visual-only (lipreading) trials---plus 10 silent control trials. Data were acquired with a CW NIRScout device (NIRx, Berlin, Germany) with 16 LED sources (760/850~nm) and photodiode detectors, sampled at 3.9~Hz. The montage comprised 44 long channels (approximately 30~mm) over the left inferior frontal gyrus, both superior temporal regions and the occipital lobe, and 8 short-separation channels (8~mm). The optodes were placed on positions of the 10--10 system without digitization. The files store the NIRx ICBM152 template positions of these sites in CapTrak coordinates; we converted them to MNI space as described in Section~\ref{CS}. The dataset is publicly available on the Open Science Framework~\cite{shader2021usedata}.

\paragraph{Relative clause processing in Turkish.}
This dataset contains recordings from native Turkish speakers who answered yes/no questions about 24 sentences, half containing a subject-relative clause and half an object-relative clause~\cite{Mutlu2023}. Fourteen participants were recorded; the file of one participant was damaged and is not included, resulting in 13 recordings. Data were acquired with a CW NIRSport device (NIRx, Berlin, Germany) with 8 LED sources (760/850~nm) and 7 detectors, forming 22 channels with a nominal SDS of 30~mm over the prefrontal cortex, sampled at 7.81~Hz. The files store optode positions identical for all participants, in an undocumented coordinate frame without anatomical landmarks. The probe figure of the study is adapted from an earlier study of the same group with the same device and number of channels~\cite{mutlu2020functional}, which reports the MNI coordinates of its 22 channels obtained with NIRS\_SPM. The channel midpoints stored in the files match these coordinates, and we therefore used them as channel positions. The dataset is publicly available on figshare~\cite{mutlu2023turkishdata}.

\paragraph{Covert and overt language production.}
This dataset contains recordings from 33 right-handed native English-speaking adults aged 20--44 years~\cite{Schroeder2023}. Participants silently read single nouns, covertly generated a verb associated with each noun, or generated it aloud. Each run lasted approximately 3.6~min, and some participants completed several sessions, resulting in 579 recordings. Data were acquired with a custom-built CW HD-DOT system (Washington University in St.~Louis) with 96 LED sources (750/850~nm) and 92 avalanche photodiode detectors, sampled at 10~Hz. The files contain 1842 source--detector pairs with first- to fourth-nearest-neighbor separations of 13, 30, 39 and 47~mm. As for the other high-density datasets, we kept the 856 pairs with SDS $\leq$ 35~mm (Appendix~\ref{app:resting}). The optode positions are identical for all participants. They are stored as voxel coordinates of the MNI152 template volume used by the NeuroDOT toolbox, and we converted them to MNI space with the toolbox function \texttt{change\_space\_coords}. The dataset is publicly available on NITRC~\cite{schroeder2023mappingdata}.

\subsection{Motor}\label{app:motor}

\paragraph{Automatic and non-automatic finger and leg movements.}
This dataset contains recordings from 24 healthy right-handed adults~\cite{Cockx2023}. Participants tapped a 12-digit sequence with the fingers of the right hand on a numeric keypad, or stepped it with the right foot on four floor pads. One sequence was practiced daily for 7 days and the other was learned on the day of the experiment. Data were acquired with two CW Brite24 devices (Artinis Medical Systems, Elst, the Netherlands) with LED sources (760/850~nm) and photodiode detectors, sampled at 50~Hz. The montage comprised 24 long channels (30~mm) and 12 short-separation channels (10~mm) over the left primary motor and premotor cortices and the bilateral dorsolateral prefrontal and posterior parietal cortices. Optode positions, together with the nasion, inion, preauricular points and vertex, were digitized for each participant from a 3D scan of the head. They were registered to MNI space with \texttt{mne.coreg.Coregistration} (Section~\ref{CS}). One participant excluded by the authors was removed. The dataset is publicly available on the Donders Repository~\cite{cockx2023fingerlegdata}.

\paragraph{Ball squeezing and resting state with a whole-head high-density probe.}
This dataset contains recordings from 17 healthy adults~\cite{carlton2026surface}. Each participant performed three runs of a ball-squeezing task. Data were acquired with a CW NinjaNIRS2022 system (Boston University, Boston, MA, USA). The whole-head high-density probe has 56 sources and 144 detectors (760/850~nm). The ball-squeezing runs were sampled at 4.32~Hz. Fourteen resting-state runs were sampled at 8.99~Hz and two at 4.32~Hz. Most resting-state files contain 567 source--detector pairs with SDS of 16--42~mm. The ball-squeezing files and one resting-state file contain 1220 pairs with SDS of up to 60~mm. We kept the pairs with SDS $\leq$ 35~mm, which gives 510--511 channels per recording. The optode positions are the same template positions as in the cocktail-party dataset (Appendix~\ref{app:speech}), and we moved them to MNI space with the same matched-point coregistration of the 10--5 landmarks. The dataset is publicly available on OpenNeuro~\cite{carlton2026ballsqueezedata}.

\paragraph{Individual finger tapping.}
This dataset contains recordings from 25 right-handed healthy adults who tapped the fingers of the right hand one at a time, in 10-s blocks from the thumb to the little finger, resulting in 69 runs of approximately 6~min~\cite{khan2026open}. Data were acquired with a CW NIRScout device (NIRx, Berlin, Germany; one participant with a NIRSport2) with 16 sources (760/850~nm) and 16 detectors forming 48 channels over the bilateral motor cortex, with an inter-optode distance of 30~mm, sampled at 3.91~Hz (10.17~Hz for the NIRSport2 recording). The shared files contain the raw intensities together with derived optical density and hemoglobin signals; only the raw intensities were used. Optode positions are reported as 10--10 labels and were assigned the corresponding MNI coordinates of the NIRx ICBM152 template. The dataset is available on figshare~\cite{khan2025fingerdata}.

\paragraph{Left- and right-hand finger tapping.}
This dataset contains recordings from 10 healthy right-handed adults who performed left- and right-hand finger tapping~\cite{mesquita2025freshmotor}. No associated publication was available. Each participant completed four runs, one for each combination of hand and block duration: 30 blocks of 2~s or 25 blocks of 3~s, separated by rest periods of 10--20~s. Data were acquired at the University of Campinas with a CW NIRScout device (NIRx, Berlin, Germany) at 760 and 850~nm, sampled at 8.93~Hz, with 14 sources and 32 detectors forming 64 long channels (approximately 30~mm) and 4 short channels (approximately 8~mm) over the sensorimotor cortex of both hemispheres. The authors digitized the optode positions of each participant, but the released files store the NIRx ICBM152 template positions in CapTrak coordinates, identical for all participants. We converted them back to MNI space as described in Section~\ref{CS}. The dataset is available on OpenNeuro~\cite{mesquita2025freshmotor}.

\paragraph{Unilateral finger tapping.}
This dataset contains recordings from 5 adults who performed a finger tapping task~\cite{luke2021tappingdata}. In each trial, participants tapped the left or right thumb to the fingers for 5~s, or rested in a control condition, following an auditory cue. Each participant was recorded once for 40--50~min. Data were acquired with a CW NIRx device (760/850~nm) with 8 sources and 16 detectors, sampled at 7.81~Hz. The montage comprised 20 long channels and 8 short-separation channels (approximately 7~mm) over the motor cortex of both hemispheres. Optode positions are a template stored in the CapTrak coordinate system, identical for all participants. They were transformed to MNI space with \texttt{mne.head\_to\_mni} (Section~\ref{CS}). The dataset is publicly available on Zenodo~\cite{luke2021tappingdata}.

\paragraph{Motor imagery BCI training after stroke.}
This dataset contains 237 sessions of motor-imagery brain--computer interface (BCI) training from 15 stroke patients with unilateral cortical lesions, who completed 7--24 sessions each over 7--15 days~\cite{isaev2024multiple}. The data were acquired with a CW NIRScout device (NIRx, Berlin, Germany) with 28 channels (14 sources, 760/850~nm, and 8 detectors; source--detector separation of about 30~mm) over the motor cortices of both hemispheres, sampled at 15.6~Hz. The sources and detectors were placed at 10--10 positions, and the files store the MNI coordinates of these positions on the NIRx ICBM152 template, which we used as given. The dataset is available from NITRC~\cite{isaev2024strokedata}.

\paragraph{Motor imagery and mental arithmetic.}
This dataset contains simultaneous EEG and fNIRS recordings from 29 healthy adults who performed left- and right-hand motor imagery and mental arithmetic versus rest, in six alternating sessions of approximately 12~min each~\cite{Shin2017}. Data were acquired with a CW NIRScout device (NIRx, Berlin, Germany) with 14 sources (760/850~nm) and 16 detectors forming 36 channels over the prefrontal, motor and occipital cortices, with an inter-optode distance of 30~mm, and are provided at 10~Hz. All channels were retained. Each optode is labeled with its position in the 10--5 system and was assigned the corresponding MNI coordinates of the NIRx ICBM152 template. The dataset is publicly available from Technische Universit\"at Berlin~\cite{Shin2017}.

\paragraph{Motor imagery of multiple upper-limb joints.}
This dataset contains simultaneous EEG and fNIRS recordings from 18 right-handed healthy adults who imagined eight movements of the right hand, wrist, elbow and shoulder, in one session of eight blocks of 40 trials each~\cite{yi2025multi}; only the fNIRS recordings were used. Data were acquired with a CW NIRScout device (NIRx, Berlin, Germany) with 8 sources (760/850~nm) and 8 detectors arranged on the 10--5 system over the left sensorimotor cortex, forming 24 channels with SDS of 32--42~mm, sampled at 7.81~Hz, resulting in 141 recordings of 13--27~min (the first four blocks of one participant were stored in a single recording); all channels were retained. Optode positions are provided as template positions in the CapTrak coordinate system and were transformed to MNI space with \texttt{mne.head\_to\_mni} (Section~\ref{CS}). The dataset is publicly available on figshare~\cite{yi2025finemidata}.

\paragraph{Directional motor imagery of the right arm.}
This dataset contains simultaneous EEG and fNIRS recordings from 30 healthy adults who imagined moving the right arm in four directions: horizontally from left to right, vertically from top to bottom, and along the two diagonals~\cite{feng2026multimodal}. The fNIRS data were acquired with a CW LABNIRS device (Shimadzu, Kyoto, Japan) with 51 channels (18 sources, 780/805/830~nm, and 18 detectors; 30-mm source--detector separation) over the sensorimotor and parietal cortices of both hemispheres, sampled at 47.62~Hz. As for the online face-gaze dataset (Appendix~\ref{app:social}), we used the hemoglobin changes computed by the Shimadzu software, expressed as concentration multiplied by optical path length, and converted them to concentrations with a 30-mm separation and a differential pathlength factor of 6. Each optode lies midway between two electrodes of the authors' 10--20 cap layout, at a 10--5 position, and we assigned it the MNI coordinates of that position on the NIRx ICBM152 template. The dataset is available from the Science Data Bank~\cite{feng2026minddata}.

\paragraph{Observation and motor imagery of balance tasks.}
This dataset contains recordings from 13 healthy adults who observed and/or imagined static and dynamic standing balance tasks~\cite{almulla2022observation}. Each participant completed three experiments in a fixed order: action observation of videos, action observation combined with motor imagery, and motor imagery guided by audio instructions. Data were acquired with a CW NIRSport2 tandem device (NIRx, Berlin, Germany) with 16 LED sources (760/850~nm) and 15 detectors, forming 40 channels with SDS of 29--31~mm (nominal 30~mm) over the prefrontal cortex and the left and right motor cortices, sampled at 3.47~Hz. The authors recorded 17 participants and excluded 4 for poor data quality. Samples flagged as saturated by the acquisition software are stored as missing values and were handled with the procedure of Section~\ref{sec:processing}; this removed one channel of one recording, which was saturated for most of the recording. Every optode lies on a 10--20 position of the probe file, identical for all participants, and was assigned the MNI position of its label on the NIRx ICBM152 template. The modified Beer--Lambert law used the nominal SDS of 30~mm. The dataset is publicly available on figshare~\cite{almulla2022balancedata}.

\paragraph{Freezing of gait in Parkinson's disease.}
This dataset contains recordings from 22 people with Parkinson's disease who reported daily freezing of gait and 22 controls~\cite{Cockx2024}. Participants walked up and down a corridor, passed through a 60-cm-wide doorway, made 180$^\circ$ turns, and stopped and started walking on instruction. Data were acquired with two CW Brite24 devices (Artinis Medical Systems, Elst, the Netherlands) with LED sources (760/850~nm) and photodiode detectors, sampled at 50~Hz and resampled to 60~Hz by the authors. The montage comprised 32 long channels (30~mm) and 16 short-separation channels (10~mm) over the primary motor, premotor, supplementary motor, prefrontal and posterior parietal cortices. The files contain optical density expressed in base-10 logarithm, which we multiplied by $\ln 10$ to obtain the natural-logarithm optical density used by the modified Beer--Lambert law. Optode positions, together with the nasion, inion, preauricular points and vertex, were digitized for each participant from a 3D scan of the head. The SNIRF files label these positions as centimeters, but the values are in millimeters, as stated in the accompanying BIDS files. We registered them to MNI space with \texttt{mne.coreg.Coregistration}, as for the finger and leg movement dataset described above. Channels marked as bad by the authors in each run were removed. The dataset is publicly available on the Donders Repository~\cite{cockx2023freezingdata}.

\paragraph{Treadmill walking.}
This dataset contains recordings from 20 healthy young adults who walked on a treadmill~\cite{minhas2024enhancing}. After a 30-s initial rest, participants completed ten trials of 20~s of walking followed by 20~s of standing rest, and the session ended with a 30-s final rest, resulting in a single recording of 7.4--7.8~min per participant. Data were acquired with a CW NIRSport2 device (NIRx, Berlin, Germany) with 8 LED sources (760/850~nm) and 8 detectors, forming 20 channels with a nominal SDS of 30~mm over the left and right motor cortices, sampled at 10.17~Hz. Optode positions are the NIRx ICBM152 template positions stored in the files, identical for all participants, and were used as stored. The dataset is publicly available on figshare~\cite{minhas2026walkingdata}.

\paragraph{Laparoscopic surgical training.}
This dataset contains recordings from four studies in which participants performed the pattern-cutting task of the Fundamentals of Laparoscopic Surgery program in a physical box trainer, cutting a marked circle out of gauze with laparoscopic tools~\cite{kamat2023assessment}. The studies compared expert and novice surgeons (14 participants), followed the learning curve of surgical residents over 12 training days (13 participants), and examined the effect of transcranial electrical stimulation on skill acquisition in a single session (12 participants) and over 12 training days with a follow-up (21 participants). The first two studies used a CW6 device (TechEn, Milford, MA, USA) operating at 690 and 830~nm, with 8 sources and 16 detectors forming 25 channels with separations of 30--40~mm, plus 8 short-separation detectors (about 8~mm), one next to each source, sampled at 20--25~Hz. The stimulation studies used a CW NIRScout device (NIRx, Berlin, Germany) operating at 760 and 850~nm, with the same montage and three additional detectors over the left and right primary motor cortices and the supplementary motor area, giving 28 long channels, sampled at 7.81~Hz. The montage covered the prefrontal cortex, the primary motor cortex and the supplementary motor area; we kept the short channels. Only the files of the stimulation studies provide a 3D layout of the probe, in an undocumented coordinate frame; we converted it to MNI space with the similarity transform that best fits its sources and long detectors to the nearest 10--5 positions of the NIRx ICBM152 template, and assigned the same positions to the identical montage of the first two studies. The dataset is available on figshare~\cite{kamat2023astaunddata}.

\paragraph{Laparoscopic training and resting state across a 24-h medical shift.}
This dataset contains recordings from 30 resident physicians~\cite{torrescuevas2025resident}. Each participant was recorded at the beginning, middle and end of a 24-h shift (0, 12 and 24~h). At each time point, the participant performed a 5-min laparoscopic training task, the folding of an origami box in a box trainer, and a 6-min eyes-closed resting-state run. This gives 180 recordings. Data were acquired with a CW Brite MKII device (Artinis Medical Systems, Elst, the Netherlands) with 10 dual-wavelength sources (757/843~nm) and 8 detectors over the prefrontal cortex, sampled at 25~Hz (75~Hz in 21 recordings). The probe forms 22 channels with an SDS of 30~mm and 2 short-separation channels. The authors provide one MNI coordinate per channel, which we used as given. We excluded the 2 short-separation channels, which have no coordinates, and one channel whose coordinates duplicate those of another channel 16~mm away on the probe. This leaves 21 channels per recording. The dataset is available on Zenodo~\cite{torrescuevas2025resident}.

\paragraph{Exercise-induced hypoalgesia.}
This dataset contains recordings from 22 endurance athletes and 20 non-athletes who performed pressure pain tests before and after high-intensity interval training on a cycle ergometer~\cite{geisler2026profound}. Data were acquired with a fiberless CW NIRSport2 device (NIRx, Berlin, Germany) with 16 dual-wavelength LED sources (760/850~nm) and 16 silicon photodiode detectors, one of which was split into 8 short-separation detectors, forming 48 long channels at 30~mm and 8 short channels at 8~mm over the prefrontal, sensorimotor and posterior parietal cortices, sampled at 7.6~Hz. The dataset is available on OpenNeuro~\cite{geisler2025athletedata}.

\subsection{Cognitive}\label{app:cognitive}

\paragraph{Visuomotor target identification.}
This dataset contains simultaneous EEG, fNIRS, eye-tracking, ECG and joystick recordings from 50 healthy adults who searched for a target aircraft among distractors and pointed to its location with a joystick~\cite{zhang2025visuomotor}. The fNIRS data were acquired with a CW ETG-4000 device (Hitachi Medical, Japan) with 44 channels (16 sources, 695/830~nm, and 14 detectors; 30-mm source--detector separation) arranged in two $3 \times 5$ probe sets over the frontal and parietal cortices, each with 8 sources and 7 detectors forming 22 channels, sampled at 10~Hz. The two sets were recorded simultaneously but stored in separate files, which we merged into a single 44-channel recording per participant. We used the channel MNI coordinates reported in the original paper, obtained by co-registration with structural MRI. For each participant, the files mark the channels flagged as exceptions by the system during setup, which show no cardiac pulsation; we discarded these channels. Samples with zero or negative intensity were handled with the procedure of Section~\ref{sec:processing}. The dataset is available on the Open Science Framework~\cite{zhao2024visuomotordata}.

\paragraph{Attention tasks after brain injury.}
This dataset contains recordings from patients with brain injury and from healthy middle-aged and older adults who performed four attention tasks~\cite{Ding2025}: the digit cancellation test, the paced auditory serial addition test, the Stroop color-word test, and the trail making test. Data were acquired with a CW NirSmart device (Danyang Huichuang Medical Equipment, Danyang, China) with 24 laser sources (730/850~nm) and 16 detectors, sampled at 11~Hz. The optodes formed 48 channels over the frontal, temporal, parietal and occipital lobes, with an SDS of 30~mm for all channels in the probe design stored in the files. The dataset provides the MNI coordinates of each channel, obtained with the standalone spatial registration of NIRS-SPM, and these coordinates were used as given. The dataset is publicly available on Mendeley Data~\cite{ding2025attentiondata}.

\paragraph{Post-error responses in multitasking.}
This dataset contains recordings from 31 adults who performed a multitasking paradigm designed to study responses after errors~\cite{zhu2025multitaskdata}. Each participant was recorded once for 37--58~min. Data were acquired with a CW NIRSport device (NIRx, Berlin, Germany) with 16 sources (760/850~nm) and 16 detectors at standard 10--5 positions, forming 39 channels with SDS of 30--42~mm over the bilateral dorsolateral prefrontal, frontopolar and orbitofrontal cortices and the angular gyrus, sampled at 4.36~Hz. Each optode was assigned the MNI position of its 10--5 label on the NIRx ICBM152 template, identical for all participants. The dataset is publicly available on the Science Data Bank~\cite{zhu2025multitaskdata}.

\paragraph{Word--color Stroop with high-density and sparse arrays.}
This dataset contains recordings from 17 healthy adults who performed a word--color Stroop task twice in one session, once with each of two probe arrays~\cite{anderson2025high}. In congruent blocks, participants chose which of two words matched its font color; in incongruent blocks, they chose which of three words named the font color of a fourth word. The data were acquired with cascaded CW NIRSport2 devices (NIRx, Berlin, Germany; two for the sparse array and four for the high-density array) with dual-tip LED sources (760/850~nm) and silicon photodiode detectors. The sparse array followed a conventional grid of 17 sources and 24 detectors, giving 52 channels with a 30-mm source--detector separation and 8 short-separation channels of about 8~mm, sampled at 24.41~Hz. The high-density array followed a hexagonal pattern of 25 sources and 66 detectors, giving 112 overlapping channels at about 19~mm, 94 at about 33~mm and 8 short-separation channels, sampled at 17.44~Hz. The files provide the positions of both arrays on the NIRx ICBM152 template in MNI coordinates, which we used as given. The datasets are available on OpenNeuro~\cite{anderson2025stroophddata}.

\paragraph{Automated driving.}
This dataset contains recordings from 20 adults who sat as passengers in a driving simulator and watched 12 simulated automated drives of about 13~min each, including cut-ins, emergency braking and pedestrians crossing the road~\cite{zhang2024fnirs}. Data were acquired with a CW OctaMon device (Artinis Medical Systems, Elst, the Netherlands) with 8 channels (8 LED sources, 757/844~nm, and 2 silicon photodiode detectors; 35-mm source--detector separation) over the prefrontal cortex, sampled at 50~Hz. The files store only a 2D probe layout with mislabeled units; we corrected the units for the Beer--Lambert conversion and assigned each channel the MNI coordinates reported by the authors. The dataset is available on OpenNeuro~\cite{zhang2024drivingdata}.

\paragraph{Cognitive and motor tasks.}
This dataset contains recordings from 30 healthy right-handed adults who performed seven 3-min tasks: mental arithmetic, auditory 2-back, motor imagery, passive and active arm movements, a combined 2-back arithmetic task, and the same combined task with active arm movements~\cite{ajra2026open}. Data were acquired with two CW devices (Artinis Medical Systems, Elst, the Netherlands): an OctaMon over the prefrontal cortex and an OctaMon+ over the left and right sensorimotor cortex, each with 8 LED sources (760/850~nm) and 2 silicon photodiode detectors forming 8 channels, with source--detector separations of 35 and 30~mm, respectively. The 16 channels were sampled at 10~Hz (50~Hz for the last ten participants). The optode positions stored in the files are the NIRx ICBM152 template positions of 10--5 locations, but in 67 of the 74 active-movement recordings the sensorimotor optodes carry the prefrontal coordinates; we therefore assigned each optode the MNI coordinates of its 10--5 location in every recording. The dataset is available on OpenNeuro~\cite{ajra2026cmx7}.

\paragraph{Mental workload during simulated flights.}
This dataset contains simultaneous EEG and fNIRS recordings from 35 young adults without flight experience who flew an Airbus A321 flight simulator while performing an auditory n-back task (0- to 3-back) and an altitude-monitoring task, in eight blocks of approximately 3~min separated by 2-min rest periods, with one recording of 43--51~min per participant~\cite{Hamann2022}. Data were acquired with a CW NIRSport2 device (NIRx, Berlin, Germany) with 8 sources (760/850~nm) and 7 detectors forming 17 channels with SDS of 26--39~mm over the bilateral prefrontal cortex, plus 8 short-separation channels, sampled at 10.17~Hz. All channels were retained. The optodes were placed on positions of the 10--5 system, and we assigned each optode the MNI coordinates of its position on the NIRx ICBM152 template. The dataset is publicly available on PsychArchives~\cite{hamann2021flightdata}.

\paragraph{Mental workload of airline pilots during simulated flights.}
This dataset contains fNIRS recordings, acquired together with EEG, eye tracking and peripheral physiological measures, from 14 commercial airline pilots who flew a simplified cruise flight in an Airbus A320 flight simulator while performing an auditory n-back task (0- to 3-back) and an altitude-monitoring task in eight blocks of approximately 5~min, with one recording of 65--78~min per participant~\cite{hamann2026examining}. The two participants that the authors excluded for sleepiness are not included in the shared data. Data were acquired with a CW NIRSport2 device (NIRx, Berlin, Germany) with 8 sources (760/850~nm) and 7 detectors forming 17 channels with SDS of 26--39~mm over the bilateral prefrontal cortex, plus 8 short-separation channels with SDS of 7--8~mm, sampled at 10.17~Hz. The montage is the same as in the simulated-flight dataset from the same group described above. Optode positions are provided on the NIRx ICBM152 template head and were used as provided. The dataset is publicly available on the Open Science Framework~\cite{hamann2025pilotdata}.

\paragraph{Working memory, discrimination and word generation.}
This dataset contains simultaneous EEG and fNIRS recordings from 26 right-handed healthy adults who performed three cognitive tasks, each in three sessions~\cite{shin2018simultaneous}. The fNIRS data were acquired with a CW NIRScout device (NIRx, Berlin, Germany) with 36 channels (16 sources, 760/850~nm, and 16 detectors; 30-mm source--detector separation) over the prefrontal (16 channels), motor (8), parietal (8) and occipital (4) cortices, sampled at 10.4~Hz. The dataset provides no measured optode positions, but the authors give the 10--5 position on which each channel is centered; we assigned each channel the MNI coordinates of that position on the NIRx ICBM152 template. The raw data are available on the authors' website at Technische Universit\"at Berlin~\cite{shin2017eegnirsdata}.

\paragraph{Symmetry in visual working memory.}
This dataset contains recordings from 37 healthy right-handed adults who performed a visual working-memory task~\cite{Sztuka2025}. After 5~min of eyes-closed rest, participants memorized the locations of two or four abstract stimuli, which were either symmetrical or asymmetrical, and then indicated the location of a probe stimulus. The task comprised 288 trials, resulting in a single recording of 47--57~min per participant. Data were acquired with two cascaded CW NIRSport2 devices (NIRx, Berlin, Germany) with dual-wavelength LED sources (760/850~nm) and avalanche photodiode detectors, sampled at 5.09~Hz. The whole-head montage comprised 32 sources and 32 detectors, forming 100 long channels (30~mm) and 16 short-separation channels (8~mm). Optode positions are the template positions of the montage designed with NIRSite (NIRx), stored in the CapTrak coordinate system and identical for all participants. They were transformed to MNI space with \texttt{mne.head\_to\_mni} (Section~\ref{CS}). Eleven participants excluded by the authors because of poor signal quality were removed, resulting in 26 recordings. The dataset is publicly available on the Open Science Framework~\cite{sztuka2024symmetrydata}.

\paragraph{Working memory load during an n-back task.}
This dataset contains recordings from healthy young adults who performed an $n$-back task with letters presented for 250~ms at intervals of 1850--2050~ms~\cite{Csipo2021}. fNIRS and transcranial Doppler sonography were recorded on separate visits, resulting in one fNIRS recording of about 10~min per participant. Data were acquired with a CW NIRScout device (NIRx, Berlin, Germany) with 16 LED sources (760/850~nm) and 16 detectors on a 128-port EasyCap, sampled at 3.91~Hz. The optodes formed 48 channels with an average SDS of 30~mm over the prefrontal cortex and the medial motor and somatosensory cortices. The dataset description gives the 10--5 position of each optode, and each optode was assigned the MNI coordinates of that position on the NIRx ICBM152 template. The authors excluded two of the 14 participants from the fNIRS analysis for excessive noise. The dataset is publicly available on PhysioNet~\cite{mukli2021nbackdata}.

\paragraph{Judgment of online rumors with fact-checking labels.}
This dataset contains recordings from 42 healthy adults who judged the authenticity of online news items~\cite{Cheng2026}. In a block design, participants read 16 true or false items for 5~s each and indicated whether each was true or false. In the first block, the items carried no source label. In the second block, they were labeled as fact-checked by an authoritative (People's Daily) or a non-authoritative (Jiupai News) source. Each item was presented five times, resulting in a single recording of 24--28~min per participant. Data were acquired with a CW NIRSport2 device (NIRx, Berlin, Germany) with 8 LED sources (760/850~nm) and 8 detectors, forming 23 channels with a nominal SDS of 30~mm over the bilateral prefrontal cortex, sampled at 10.17~Hz. All channels were retained. Optode positions are the NIRx ICBM152 template positions stored in the files, identical for all participants, and were used as stored. The dataset also contains duplicate copies of four sessions, which were read only once. The dataset is publicly available on the Science Data Bank~\cite{ding2026rumordata}.

\paragraph{Real use of a cosmetic product.}
This dataset contains recordings from 30 women who applied six foundations to their face and then rated how much they would be willing to pay for each, in one session of 10--14~min~\cite{KawabataDuncan2019}. Data were acquired with a CW ETG-4000 device (Hitachi Medical Corporation, Kashiwa, Japan) with a $3\times11$ holder of 17 illuminators (695/830~nm) and 16 detectors forming 52 channels with an inter-optode distance of 30~mm over the frontal and temporal cortices, sampled at 10~Hz. The shared files contain the raw intensities; channels marked as exceptions by the device were excluded, leaving 41--52 channels per recording. The optode positions, together with the nasion and preauricular points, were digitized for each participant and registered to MNI space with \texttt{mne.coreg.Coregistration} (Section~\ref{CS}). The dataset is publicly available on figshare~\cite{kawabataduncan2018cosmeticdata}.

\paragraph{Source-authority cues and perceived artificial intelligence capability.}
This dataset contains task recordings from 72 adults, divided into three groups of 24, from a study on how source-authority cues shape the perceived capability of artificial intelligence~\cite{ding2026wired}. Each participant was recorded once for 3--6~min. Data were acquired with a CW NIRSport2 device (NIRx, Berlin, Germany) using 8 LED sources (760/850~nm) and 8 detectors, forming 23 channels with a nominal SDS of 30~mm over the bilateral prefrontal cortex, sampled at 10.17~Hz. Optode positions are the NIRx ICBM152 template positions stored in the files, identical for all participants, and were used as stored. The dataset is publicly available on the Science Data Bank~\cite{ding2026aipersuasiondata}.

\subsection{Visual}\label{app:visual}

\paragraph{Animated movie viewing and functional localizers.}
This dataset contains recordings from 5 adults acquired during the validation of a high-density diffuse optical tomography (HD-DOT) system designed for imaging awake young children~\cite{tripathy2024mapping}. Participants completed functional localizer tasks (a rotating flickering checkerboard wedge, passive listening to spoken word lists and left- and right-hand finger tapping), and four of them viewed 5--6-min clips of animated children's movies over multiple sessions, resulting in 189 runs of 3--7~min across 1--8 sessions per participant, 142 of them movie viewing. Data were acquired with a custom-built CW HD-DOT system (Washington University in St.~Louis, MO, USA) with 128 dual-wavelength laser-diode sources (685/830~nm) and 125 avalanche photodiode detectors covering the occipital, left and right temporal and dorsal regions of the head, sampled at 10.42~Hz. The files contain 3375--3515 source--detector pairs with SDS of 9--60~mm. As for the other high-density datasets, we retained the pairs with SDS $\leq$ 35~mm (Appendix~\ref{app:resting}), yielding 1434--1465 channels per recording, of which 426--429 are short-separation channels (SDS $<$ 12~mm). Optode positions were registered to MNI space with \texttt{mne.coreg.Coregistration}, as for the Alzheimer's disease resting-state dataset (Appendix~\ref{app:resting}). The dataset is publicly available on OpenNeuro~\cite{fogarty2026hddotdata}.

\paragraph{Whole-head very-high-density functional localizers and movie viewing.}
This dataset contains recordings from 8 healthy adults acquired during the validation of a whole-head very-high-density diffuse optical tomography (VHD-DOT) system~\cite{fogarty2025functional}. In two sessions, participants completed functional localizer tasks (passive listening to spoken word lists, covert verb generation, left and right flickering checkerboard wedges, and left- and right-hand finger tapping) and viewed a 10-min clip of the movie \textit{The Good, the Bad, and the Ugly} twice per session, resulting in 88 runs of 3--11~min. Data were acquired with a custom-built CW VHD-DOT system (Washington University in St.~Louis, MO, USA) with 255 dual-wavelength laser-diode sources (685/830~nm) and 252 avalanche photodiode detectors on a 9.75-mm grid over four cap panels covering the left, right, posterior and dorsal regions of the head, sampled at 7.81~Hz. The files contain 9543--10759 source--detector pairs with SDS of 8.5--60~mm; we retained the pairs with SDS $\leq$ 35~mm, yielding 3325--3652 channels per recording, of which 906--909 are short-separation channels (SDS $<$ 12~mm). Optode positions were placed on each participant's MRI-based head model, guided by session photographs and word-hearing fMRI maps. They were registered to MNI space with \texttt{mne.coreg.Coregistration}, as for the Alzheimer's disease resting-state dataset (Appendix~\ref{app:resting}). The dataset is publicly available on OpenNeuro~\cite{fogarty2026vhddotdata}.

\paragraph{Visual semantics of naturalistic movies with very-high-density diffuse optical tomography.}
This dataset contains recordings from 6 healthy adults who passively viewed silent natural movie clips~\cite{Fehner2026}. Each participant completed three imaging sessions on the same day. Each session included four 10-min runs of unique training movies, three 10-min runs of 1-min test clips repeated in randomized order, a visual checkerboard localizer and an auditory word-list localizer, resulting in 178 runs of 3.5--10.3~min. Data were acquired with the same custom-built CW VHD-DOT system (Washington University in St.~Louis, MO, USA) as the whole-head functional localizer dataset described above, with 255 dual-wavelength laser-diode sources (685/830~nm) and 252 avalanche photodiode detectors on a 9.75-mm grid, sampled at 7.81~Hz. As for the other high-density datasets, we retained the pairs with SDS $\leq$ 35~mm (Appendix~\ref{app:resting}), yielding 3406--3955 channels per recording, of which 905--927 are short-separation channels (SDS $<$ 12~mm). Optode positions were placed on each participant's MRI-based head model, aligned to session photographs and refined with the localizer maps. They are identical across the three sessions of a participant. They were registered to MNI space with \texttt{mne.coreg.Coregistration}, as for the whole-head functional localizer dataset. The dataset is publicly available on OpenNeuro~\cite{fehner2026semanticdata}.

\paragraph{Visual task in Alzheimer's disease and mild cognitive impairment.}
This dataset comes from the same study and participants as the Alzheimer's disease resting-state dataset (Appendix~\ref{app:resting})~\cite{Butters2026}. It contains one visual-task recording per participant, of 8--11~min. Data were acquired with the same CW LUMO system (Gowerlabs, London, UK), here with 6 modules over the occipital cortex. The array has 18 dual-wavelength LED sources (735/850~nm) and 24 photodiode detectors, which form 432 source--detector pairs, and was sampled at 5~Hz. As for that dataset, we kept the pairs with SDS $\leq$ 35~mm. This gives 143--191 channels per recording, of which 54--55 are short-separation channels (SDS $<$ 12~mm). One file is truncated in the public upload and cannot be read, which leaves 63 recordings. The optode positions were digitized on each participant's head together with the nasion, inion, preauricular points and vertex. We registered them to MNI space with \texttt{mne.coreg.Coregistration}, as for the resting-state dataset. The dataset is publicly available on OpenNeuro~\cite{butters2026cambridgevisualdata}.

\paragraph{Visual stimulation and breath-holding during calibrated fMRI.}
This dataset contains fNIRS recordings acquired concurrently with fMRI scanning from 13 healthy adults~\cite{rocco2026comparison}. Each participant performed a breath-hold task followed by a visual stimulation task. Data were acquired with a CW Borealis device (NIRx, Berlin, Germany) with MR-compatible laser sources (785/830~nm) and avalanche photodiode detectors, sampled at 14.8~Hz. Fifteen sources and 17 detectors were placed bilaterally over the occipital lobe. They formed 54 long channels with SDS of 21--42~mm and 8 short-separation channels with SDS of 8~mm. The files list all 114 source--detector pairs. We retained the 62 pairs of the montage (SDS $\leq$ 45~mm), as in the original study. Optode positions are the NIRx ICBM152 template positions stored in the files, identical for all participants, and were used as stored. The authors excluded three participants for insufficient fNIRS data quality, defined as fewer than 60\% of channels passing their quality check. We removed the same participants, resulting in 20 recordings from 10 participants. The dataset is publicly available on the G-Node GIN repository~\cite{rocco2026domribraindata}.

\section{Downstream datasets}\label{sec:downstream_datasets}

The six datasets held out for downstream evaluation are described below in the order of Table~\ref{tab:downstream}. Their labeling and epoching are described in Section~\ref{sub:downstream}.

\paragraph{Intracerebral hemorrhage (ICH).}
This dataset contains fNIRS recordings, acquired simultaneously with EEG, from 37 right-handed participants who imagined grasping with the left or right hand: 17 healthy adults and 20 patients with ICH~\cite{shi2025hefmi}. Data were acquired with a CW NirScan device (Danyang Huichuang Medical Equipment, Danyang, China) with 90 channels (32 laser sources, 730/808/850~nm, and 30 detectors; 30-mm source--detector separation) over the prefrontal, motor and association cortices, sampled at 11~Hz. Of the three wavelengths, we used the 730 and 850~nm signals for the modified Beer--Lambert law. We assigned each optode the MNI coordinates reported by the authors. The dataset is available on figshare~\cite{shi2025hefmidata}.

\paragraph{Multimodal speech comprehension with cochlear implants (CI status).}
This dataset contains recordings from 72 adults, 46 cochlear implant users and 26 normal-hearing controls, who listened to and watched sentences of the German Matrix Sentence Test under four conditions (speech in quiet, speech in noise, audiovisual speech and visual speech, i.e., lipreading)~\cite{balint2025fnirs}. Each participant was recorded once for 26--29~min with a CW FOIRE-3000 device (Shimadzu, Kyoto, Japan) with 39 channels (16 laser sources, 780/805/830~nm, and 16 photomultiplier detectors), three of them short-separation channels of about 15~mm, over the prefrontal, temporal and occipital cortices, sampled at 14.29~Hz. The files contain absorbance at the three wavelengths; since the modified Beer--Lambert law applied to the other datasets uses two wavelengths, we used the 780 and 830~nm signals as optical density. The authors set to missing the channels that could not be recorded because of the implant's transmitter or the cap setup (0--8 per recording), which we discarded. The probe positions were obtained from a 3D scan of each participant's head and aligned by the authors to the MNE 10--05 template in MNI space; we used them as given. The dataset is available on Dryad~\cite{balint2025speechdata}.

\paragraph{Auditory stimulation in tinnitus (Tinnitus).}
This dataset contains recordings from 63 adults, 29 with subjective tinnitus and 34 controls, acquired during 3~min of silence followed by 12-s blocks of broadband noise, with one recording of 16--20~min per participant~\cite{satish2026tinnitusdata}. There is no associated publication for this dataset. Data were acquired with a CW NIRSport2 device (NIRx, Berlin, Germany) with 24 sources (760/850~nm) and 31 detectors, forming 83 channels, 8 of them short-separation channels (SDS of 8--9~mm) and 75 long channels, sampled at 5.09~Hz. Optode positions are provided on the NIRx ICBM152 template head, and were used as provided. The dataset is publicly available on OpenNeuro~\cite{satish2026tinnitusdata}.

\paragraph{Motor recovery after stroke (Movement).}
This dataset contains recordings from 16 patients with upper-limb hemiparesis after stroke, acquired in 3--6 sessions per patient, 84 in total, during a two-week inpatient rehabilitation program~\cite{medvedeva2026multisession}. EEG and fNIRS were recorded simultaneously; the fNIRS data were acquired with two CW NIRSport devices (16$\times$16 and 8$\times$8; NIRx, Berlin, Germany) with 24 sources (760/850~nm) and 24 silicon photodiode detectors placed about 30~mm apart over the sensorimotor and adjacent areas, sampled at 13.56~Hz. The authors excluded a malfunctioning detector and its symmetric counterpart, leaving 70 channels. The files provide optical density and the MNI coordinates of the optodes' 10--20 positions on the NIRx ICBM152 template. The dataset is available on figshare~\cite{medvedeva2025strokedata}.

\paragraph{Action observation and execution (Action mode).}
This dataset contains recordings from 21 healthy adults in a task with five conditions~\cite{li2020brain}: watching a video of a right hand placing a card into an open box (simple action), watching a video of a right hand inserting a card through a narrow slot (complex action), performing the simple action, performing the complex action, and, as a control, watching a video without any hand movement. Data were acquired with a CW7 device (TechEn, Milford, MA, USA) operating at 690 and 830~nm, with 16 sources, 24 long-separation detectors and 8 short-separation detectors forming 52 long channels (30~mm) and 8 short channels (8~mm) over the inferior frontal to posterior parietal regions of both hemispheres, sampled at 25~Hz. Although the optode positions were digitized, the files store only a 2D probe layout, so we assigned each channel the MNI coordinates reported by the authors. We excluded one duplicated run and two runs whose files contain corrupted values, keeping 61 recordings. The dataset is available on OpenNeuro~\cite{li2025mirrordata}.

\paragraph{Semantic imagery of animals and tools (Semantic category).}
This dataset contains simultaneous EEG and fNIRS recordings from 12 right-handed adults who viewed images of 18 animals and 18 tools and, after each image, performed four 3-s mental tasks: silently naming the object and imagining seeing, hearing and touching it~\cite{rybavr2025simultaneous}. The fNIRS data were acquired with a CW NIRScoutXP device (NIRx, Berlin, Germany) with low-profile optodes operating at 785 and 830~nm, using one of two montages over the left hemisphere: a frontal montage with 8 sources and 4 detectors forming 14 channels, sampled at 7.81~Hz (6 participants), and a temporal--parietal montage with 7 sources and 4 detectors forming 11 channels, sampled at 8.93~Hz (the other 6 participants). We identified each optode with the 10--5 position given in the original paper and assigned it the MNI coordinates of that position on the NIRx ICBM152 template. The dataset is publicly available on OpenNeuro~\cite{rybar2025animalsdata}.

\section{Additional datasets without 3D positions}\label{sec:rejected}

The 15 datasets described below are released separately because they do not provide optode positions that can be placed in MNI space (see Table~\ref{tab:rejected}). Thirteen datasets store only a two-dimensional probe layout, and two store three-dimensional coordinates in an undocumented frame without anatomical landmarks. Their loaders apply the same processing as the corpus (Section~\ref{sec:processing}) and keep the positions exactly as provided, in the native frame of each dataset. Together they hold 832 recordings and 182~h from 348 participants.

\begin{table}[t]
\centering
\footnotesize
\setlength{\tabcolsep}{3pt}
\caption{\textbf{Additional datasets without 3D positions}. The 15 datasets released with loaders but not included in the corpus, ordered by recording time within each group.}
\label{tab:rejected}
\begin{tabular}{@{}>{\raggedright\arraybackslash}p{3.4cm}>{\raggedright\arraybackslash}p{3.4cm}>{\raggedright\arraybackslash}p{2.2cm}rr>{\raggedright\arraybackslash}p{2.2cm}@{}}
\toprule
\textbf{Dataset} & \textbf{Task} & \textbf{Population} & \textbf{Participants} & \textbf{Hours} & \textbf{Ref.} \\
\midrule
\multicolumn{6}{c}{\textit{Two-dimensional coordinates}} \\
Multi-day Stroop & Stroop, three sessions & Healthy & 55 & 47.90 & \cite{zeng2026stroop,zeng2026stroopdata} \\
Emotional pictures & Picture viewing, affective imagery & Healthy & 31 & 23.87 & \cite{spape2024nemo,spape2023nemodata} \\
Stroop neurofeedback & Stroop & Healthy & 45 & 21.15 & \cite{zeng2025stroopblockdata} \\
Parkinson's disease & Rest, finger tapping, walking & PD, healthy & 40 & 12.68 & \cite{guevara2024parkinson,guevara2023parkinsondata} \\
Morphine and placebo & Electrical pain stimuli & Healthy & 14 & 12.33 & \cite{peng2018morphine,peng2025morphinedata} \\
Audiovisual attention & Attention to one of three talkers & Healthy & 12 & 10.46 & \cite{ning2024complex,ning2025complexdata} \\
Binocular color rivalry & Dichoptic color stimuli & Healthy & 10 & 6.95 & \cite{deng2025rivalrydata} \\
Electrical and thermal pain & Pain stimuli, finger tapping & Healthy & 11 & 6.91 & \cite{yucel2015specificity,yucel2015short,yucel2025thermaldata} \\
Tinnitus resting state & Rest & Tinnitus, healthy & 18 & 5.30 & \cite{sanjuan2017tinnitus,sanjuan2017tinnitusdata} \\
Remifentanil & Surgical stimulation under anesthesia & Cardiac patients & 2 & 3.09 & \cite{karunakaran2022remifentanil,karunakaran2022remifentanildata} \\
Occipital resting state & Rest & Healthy & 14 & 1.17 & \cite{vonluhmann2020open,vonluhmann2020restingdata} \\
Motion artifacts I & Head and face movements & Healthy & 7 & 1.08 & \cite{yucel2014reducing,yucel2025motiondata} \\
Motion artifacts II & Rest with motion artifacts & Healthy & 5 & 0.52 & \cite{yucel2020motion2data} \\
\midrule
\multicolumn{6}{c}{\textit{Three-dimensional coordinates in an undocumented frame}} \\
Mindfulness and VR shooting & Rest, VR shooting competition & Athletes & 72 & 21.50 & \cite{gao2023mindfulness,gao2022mindfulnessdata} \\
Multi-distance ball squeezing & Rest, ball squeezing, motion & Healthy & 12 & 7.52 & \cite{gao2023short,zhou2026multidistancedata} \\
\midrule
Total & & & 348 & 182.43 & \\
\bottomrule
\end{tabular}
\end{table}

\paragraph{Motion artifacts with collodion-fixed optical fibers.}
This dataset contains recordings from 7 healthy adults who performed deliberate head and face movements, such as reading aloud, nodding, head twisting and eyebrow raising~\cite{yucel2014reducing}. Data were acquired with a CW6 system (TechEn, Milford, MA, USA) at 690/830~nm and 50~Hz, with 14 channels at 30~mm over both motor cortices. The files store a flat 2D probe drawing in cm without anatomical landmarks, and the paper shows the probe position only as a figure on an atlas head. The dataset is publicly available on OpenNeuro~\cite{yucel2025motiondata}.

\paragraph{Motion artifacts, dataset II.}
This dataset from the same laboratory~\cite{yucel2014reducing} contains five resting-state recordings of 3.7--7.6~min from 5 participants, released as a second set of recordings with motion artifacts~\cite{yucel2020motion2data}. No publication or documentation describes the recordings. Its files store only 2D optode positions, without 3D positions or landmarks. The dataset is publicly available on NITRC~\cite{yucel2020motion2data}.

\paragraph{Finger tapping, walking and rest in Parkinson's disease.}
This dataset contains recordings from 20 patients with Parkinson's disease and 20 controls during rest, finger tapping and walking~\cite{guevara2024parkinson}. Data were acquired with a CW Brite MKII device (Artinis Medical Systems, Elst, the Netherlands) at 757/843~nm, with 20 long channels at 30~mm and 2 short channels over both motor cortices. The paper states that the optode positions were digitized, but the files store only a flat 2D drawing without landmarks. The dataset is publicly available on Zenodo~\cite{guevara2023parkinsondata}.

\paragraph{Painful electrical stimuli under morphine and placebo.}
This dataset contains recordings from 14 healthy men who received innocuous and noxious electrical stimuli on the thumb before and after taking morphine or a placebo~\cite{peng2018morphine}. Data were acquired with a CW7 system (TechEn, Milford, MA, USA) at 690/830~nm and 25~Hz, with 24 long channels at 30~mm and 9 short channels at 8~mm. The files store a flat 2D layout with all depth coordinates equal to zero and no landmarks. The dataset is publicly available on OpenNeuro~\cite{peng2025morphinedata}.

\paragraph{Emotional pictures and affective imagery.}
This dataset contains recordings from 31 adults who viewed emotional pictures and performed affective imagery in a single session of approximately 48~min~\cite{spape2024nemo}. Data were acquired with a CW Brite24 device (Artinis Medical Systems, Elst, the Netherlands) at 756/853~nm and 50~Hz, with 24 channels at 30~mm over the frontal areas. The files store a flat grid with all depth coordinates equal to zero. The dataset is publicly available on Zenodo~\cite{spape2023nemodata}.

\paragraph{Audiovisual spatial attention with three talkers.}
This dataset contains recordings from 12 adults who attended to one of three talkers presented in videos~\cite{ning2024complex}. Data were acquired with a CW6 system (TechEn, Milford, MA, USA) at 690/830~nm and 50~Hz, with 30 long channels at 28--30~mm and 6 short channels at 8~mm. The files store only the 2D probe design from AtlasViewer. The dataset is publicly available on OpenNeuro~\cite{ning2025complexdata}.

\paragraph{Electrical and thermal pain and finger tapping.}
This dataset contains recordings from 11 healthy men during electrical stimulation, finger tapping and other tasks~\cite{yucel2015specificity,yucel2015short}. Data were acquired with a CW6 system (TechEn, Milford, MA, USA) at 690/830~nm and 50~Hz, with 36 long channels at 30~mm and 15 short channels at 8~mm over both sensorimotor cortices and the frontal cortex. The published study states that the optode positions were digitized, but the files contain only the 2D probe design without landmarks, and the papers give MNI coordinates for only three channels. The dataset is publicly available on OpenNeuro~\cite{yucel2025thermaldata}.

\paragraph{Resting-state functional connectivity in tinnitus.}
This dataset contains resting-state recordings from 10 patients with tinnitus and 8 controls~\cite{sanjuan2017tinnitus}. Data were acquired with a CW6 system (TechEn, Milford, MA, USA) at 690/830~nm and 20~Hz, with 44 channels over both hemispheres. The paper states that the optode positions were digitized, but the files contain only a flat 2D layout with all depth coordinates equal to zero. The dataset is publicly available on Harvard Dataverse~\cite{sanjuan2017tinnitusdata}.

\paragraph{Multi-day Stroop decoding.}
This dataset contains recordings from 55 adults who performed a block-design Stroop task in three sessions on different days~\cite{zeng2026stroop}. Data were acquired with a CW LABNIRS device (Shimadzu, Kyoto, Japan) at 780/805/830~nm and 25.6~Hz, with 35 channels at 30~mm over the prefrontal cortex. The files contain only hemoglobin concentrations computed by the device software. The paper states that the channel positions were digitized and registered to MNI space, but no coordinates are shared, and the probe files contain only a flat 2D layout. The dataset is publicly available on figshare~\cite{zeng2026stroopdata}.

\paragraph{Stroop task in a decoded-neurofeedback study.}
This dataset contains recordings from 45 participants who performed a block-design Chinese color--word Stroop task of 30 blocks lasting about 25~min~\cite{zeng2025stroopblockdata}. Data were acquired with a CW LABNIRS device (Shimadzu, Kyoto, Japan) with 35 channels at 25.6~Hz, and the files store only device-computed hemoglobin concentrations. We did not find a paper describing the dataset. The dataset is publicly available on the Science Data Bank~\cite{zeng2025stroopblockdata}.

\paragraph{Occipital resting state with synthetic hemodynamic responses.}
This dataset contains 5-min resting-state recordings from 14 healthy adults who fixated a cross~\cite{vonluhmann2020open}. Data were acquired with a CW6 system (TechEn, Milford, MA, USA) at 690/830~nm and 50~Hz, with 8 sources, 12 long-separation detectors at about 30~mm and 2 short-separation detectors at about 10~mm, forming 26 long and 2 short channels over the occipital lobe. The probe was designed in AtlasViewer; the files store only 2D optode positions, and the paper shows the probe only as a figure. The dataset is publicly available on NITRC~\cite{vonluhmann2020restingdata}.

\paragraph{Binocular color rivalry along red--green and yellow--blue directions.}
This dataset contains recordings during binocular color rivalry along the red--green and yellow--blue directions~\cite{deng2025rivalrydata}. Data were acquired with a CW NirSmart device (Danyang Huichuang Medical Equipment, Danyang, China) at 730/850~nm and 20~Hz, with 14 channels designed at 30~mm. The files store a flat 2D probe grid. No paper describing the dataset was found. The dataset is publicly available on figshare~\cite{deng2025rivalrydata}.

\paragraph{Nociception during cardiac ablation under general anesthesia.}
This dataset comes from a randomized trial of remifentanil (an opioid) in 41 patients aged 12--30 years who underwent catheter ablation of a cardiac arrhythmia under general anesthesia~\cite{karunakaran2022remifentanil}. Each patient received placebo, a low dose or a high dose of remifentanil, and fNIRS was recorded over the prefrontal and somatosensory cortices to measure the brain response to the surgical stimulation. Data were acquired with a CW7 system (TechEn, Milford, MA, USA) at 690/830~nm and 25~Hz, with 9 sources, 12 detectors at 30~mm and 9 short-separation detectors at 8~mm, forming 24 long and 9 short channels. Only 10 of the 225 files of the repository are public. The files store a flat 2D probe layout. The dataset is available on the Harvard Dataverse~\cite{karunakaran2022remifentanildata}.

\paragraph{Ball squeezing with a multi-distance high-density probe.}
This dataset contains recordings from 12 adults during a 5-min rest, three 8-min runs of left- and right-hand ball squeezing and a motion-artifact task~\cite{gao2023short}. Data were acquired with two cascaded CW NIRSport2 devices (NIRx, Berlin, Germany) at 760/850~nm and 8.72~Hz, with 14 sources and 32 detectors forming 100 channels at 17--41~mm. The files store 3D optode positions, but their coordinate frame is undocumented. The dataset is publicly available on OpenNeuro~\cite{zhou2026multidistancedata}.

\paragraph{Brief mindfulness meditation before a virtual reality shooting competition.}
This dataset contains recordings from university athletes during a baseline and a 30-min virtual reality shooting competition after 20~min of mindfulness meditation or mind wandering; the paper reports 76 eligible participants~\cite{gao2023mindfulness}. Data were acquired with a CW high-density LUMO system (Gowerlabs, London, UK) of 9 tiles over the prefrontal cortex, with 27 sources at 735/850~nm and 36 detectors forming 972 channels at 10--45~mm, sampled at 5~Hz. The files store 3D optode coordinates in an undocumented frame without anatomical landmarks. The dataset is publicly available on Mendeley Data~\cite{gao2022mindfulnessdata}.


\end{document}